\documentclass[final,5p,times,twocolumn,number,square,comma]{elsarticle}

\usepackage{amsmath, amssymb, amsthm, amsfonts}
\usepackage{graphicx}
\usepackage{textcomp}
\usepackage{xcolor}
\usepackage{enumerate}
\usepackage{amsmath}
\usepackage{amssymb}
\usepackage{algorithm}
\usepackage{algorithmic}
\usepackage{caption}
\usepackage{booktabs, multirow, multicol}
\usepackage{etoolbox}
\appto{\bibsetup}{\renewcommand{\url}[1]{}}
\usepackage{natbib}
\DeclareMathOperator*{\argmin}{arg\,min}

\newtheorem{myDef}{Definition}



\journal{Journal of the Franklin Institute}

\begin{document}

\begin{frontmatter}



\title{\textbf{Dy-mer}: An Explainable DNA Sequence Representation Scheme using Dictionary Learning}


\author{Zhiyuan Peng$^{\dagger}$, Naifan Zhang$^{\dagger}$, Yuanbo Tang, Yang Li$^{*}$}
\affiliation[label1]{
        organization={Shenzhen Key Laboratory of Ubiquitous Data Enabling},
        addressline={Tsinghua Shenzhen International Graduate School, Tsinghua University},
        city={Shenzhen},
        postcode={518055},
        state={Guangdong},
        country={China}
}

\begin{abstract}
DNA sequences encode critical genetic information, yet their variable length and discrete nature impede direct utilization in deep learning models. Existing DNA representation schemes convert sequences into numerical vectors but fail to capture structural features of local subsequences and often suffer from limited interpretability and poor generalization on small datasets. 
To address these limitations, we propose \textbf{Dy-mer}, an interpretable and robust DNA representation scheme based on dictionary learning. Dy-mer formulates an optimization problem in tensor format, which ensures computational efficiency in batch processing. Our scheme reconstructs DNA sequences as concatenations of dynamic-length subsequences (dymers) through a convolution operation and   simultaneously optimize a learnable dymer dictionary and sparse representations.
Our method achieves state-of-the-art performance in downstream tasks such as DNA promoter classification and motif detection.  Experiments further show that the learned dymers match known DNA motifs and  clustering using Dy-mer yields semantically meaningful phylogenetic trees. These results demonstrate that the proposed approach achieves both strong predictive performance and high interpretability, making it well suited for biological research applications.

\end{abstract}




\end{frontmatter}


\section{Introduction}
\label{}

DNA sequences contain genetic and biological information critical for organismal development and function~\cite{dna}. Typically, DNA adopts a double helix structure composed of two intertwined strands of alternating deoxyribonucleotides. Each nucleotide comprises a deoxyribose sugar, a phosphate group, and one of four nucleobases: adenine (A), cytosine (C), guanine (G), or thymine (T). Biologists have established that nucleotide permutations encode vital biological information essential for RNA transcription and protein synthesis. However, raw DNA sequences exhibit inherent length variability and non-numerical characteristics, rendering them incompatible with most data-mining models. Hence, diverse DNA representation schemes have been developed to convert raw sequences into fixed-length numerical vectors, enabling applications such as similarity analysis~\cite{JIN2017342} and classification~\cite{Sun2021iPTT2LA}.

Early studies focused solely on individual nucleotides, often mapping the four nucleobases to geometric coordinates or numerical values. For example, a 2-dimensional model~\cite{RANDIC2003202} mapped nucleotides to Cartesian axes, while others assigned specific numerical values~\cite{AFREIXO201152,950245,5189632} or frequency metrics~\cite{biomimetics8020218} to each type. However, these schemes overlooked structure-aware features of local DNA subsequences. Termed K-mers (with $K$ denoting subsequence length), these local segments encode rich semantic information about underlying structural and biological characteristics.
This insight stems from observed sequence patterns: biologists have identified recurring K-mers that consistently perform analogous functions, such as transcription factor binding sites (TFBS). These recurring patterns, referred to as motifs~\cite{motif, conseq}, play a pivotal role in DNA regulation. Thus, representing a DNA sequence as a set of K-mers is more biologically meaningful. For instance, one scheme employed Empirical Mode Decomposition to extract features from nucleotide permutations, facilitating the identification of characteristic DNA segments~\cite{BAI2011232}. Le et al.~\cite{noauthor_frontiers_nodate} used sliding windows to generate K-mers~\cite{kmer1, kmer2}, applied FastText N-Grams for vector conversion, and aggregated these vectors to represent full sequences. There are also numerous LLM-based approaches that leverage biological data to fine-tune large models, which are then utilized to embed DNA sequences by treating K-mers as tokens~\cite{llm}. 


While existing deep learning-based representation schemes exhibit promising downstream task performance, they often lack interpretable structures, relying instead on attention or gradient mechanisms of task-specific algorithms~\cite{PMID:34903170, SANTORSOLA20231} to implicitly quantify feature contributions (e.g., embedded K-mers). This impedes biologists' ability to validate and utilize derived insights.  To address these limitations, we represent each DNA sequence as the concatenation of variable length subsequences, namely \textit{dymer}s, a portmanteau of \textit{dynamic} and \textit{K-mers}, providing a  representation structure that  captures both positional and sequential information with strong interpretability.

The core challenge then becomes how to extract meaningful dymers and reconstruct sequences optimally, given the inherent complexities of DNA data. DNA sequences are high-dimensional with diverse structural patterns and high-quality datasets are often scarce due to frequent mutations (insertions, deletions, and point mutations). 
Nevertheless, DNA sequences exhibit inherent regularity and recurring patterns, which enable us to construct a finite dictionary of dymers that capture these patterns and represent each DNA sequence through sparse indexing of this dictionary. To achieve this, we are inspired by sparse representation learning ~\cite{candes2006robust, donoho2006compressed, Chu2022.11.06.515322}, which recovers underlying sparse structures from raw data under the assumption that signals can be sparsely expressed via a well-designed basis. 

The quality of the dictionary directly determines the effectiveness of sparse representations. Predefined dictionaries constructed from frequent real-world K-mers are highly sensitive to hyperparameters: over-sampling introduces severe redundancy (e.g., \textit{AAAAAA} and \textit{AAAA}), inflating the basis size and dimensionality, whereas under-sampling fails to capture the intrinsic complexity of DNA sequences. To overcome these limitations, we propose \textbf{Dy-mer}, a DNA representation framework that jointly learns an optimal dymer dictionary from training data while recovering sparse representations.

In contrast to conventional dictionary learning, \textbf{Dy-mer} encodes both positional and sequential information into a two-dimensional sparse representation and reconstructs DNA sequences through a convolutional formulation. Moreover, \textbf{Dy-mer} introduces two complementary sparsity constraints to promote dictionary compactness and representation parsimony. Guided by the Minimum Description Length (MDL) principle, these regularizers enable optimal basis selection from an overcomplete dictionary and yield more informative and robust DNA representations. To improve computational efficiency, \textbf{Dy-mer} adopts tensor-based data structures and operations to support efficient batch processing. Extensive experiments demonstrate state-of-the-art performance on downstream tasks (e.g., DNA promoter classification and DNA clustering) while significantly enhancing interpretability for DNA motif discovery.

In summary, the main contributions of our research are as follows:
\begin{itemize}
    \item We proposed a robust and interpretable DNA representation scheme based on dictionary learning, inspired by the minimum description length principle. 

    \item   We designed a 2-dimensional position-aware sparse representation structure, regarding DNA sequences as the convolutional reconstruction with learned dymers, enabling explicit encoding of motif positions and repetitions that are critical for sequence analysis.
    
    \item We demonstrate the effectiveness of our DNA representations on multiple downstream tasks. In DNA promoter classification, our method achieves at least a 14\% and 13\% improvement in classification accuracy over representative non-LLM-based and LLM-based DNA representation learning methods, respectively.
\end{itemize}

The subsequent sections of this article are structured as follows: Section 2 is the literature review of DNA representation schemes. Section 3 provides essential background knowledge necessary for our representation scheme. Section 4 introduces the explainable representation structure of our scheme. Sections 5 and 6 elaborate on our methodology based on sparse representation learning and dictionary learning, respectively. Finally, we offer conclusions and insights derived from our research in Section 7.

\section{Literature Review}

DNA representation schemes serve a pivotal role in computational biology, transforming dynamic-length DNA sequences into fixed-length representations. Different schemes are designed for specific applications, but their common objective is to extract meaningful features, such as chemical properties, composition, and permutation of nucleotides, to generate effective representations.

\subsection{Classification and Evolution of DNA Representation Schemes}

There are two main classes: graphical and numerical schemes. Graphical schemes~\cite{RANDIC2003202} aim at mapping important biochemical features to different geometrical alternatives. On the contrary, numerical schemes encode sequence information mathematically, offering effectiveness and scalability in high-dimensional and high-throughput settings. Traditionally, researchers mapped individual nucleotides into numerical values~\cite{AFREIXO201152,950245,5189632,biomimetics8020218}. Later, a method employing Empirical Mode Decomposition obtains several intrinsic mode functions to capture more complex subsequences~\cite{BAI2011232}.


The emergence of deep learning methods provides a powerful tool for DNA representation learning, especially the large language models (LLMs). FastText N-Grams~\cite{noauthor_frontiers_nodate} and word2vec~\cite{ng2017dna2vec} are applied to convert K-mers of varying lengths into DNA representations. Researchers have fine-tuned different LLMs on biological data, such as DNABert~\cite{10.1093/bioinformatics/btab083}, DNAGPT~\cite{DBLP:journals/corr/abs-2307-05628}, DNABERT2~\cite{zhou2024dnabert}, and GROVER~\cite{nature}. Meanwhile, different tokenization methods are employed to enhance the performance. HyenaDNA~\cite{10.5555/3666122.3667994} employs convolutional neural networks (CNNs) for implicit token generation, other models such as DNABERT2 and GROVER leverage byte-pair encoding to generate K-mer sets with more balanced frequency distributions.

Many researchers have leveraged DL for sparse representations learning\cite{7102696, survey1}. Researchers have proposed that natural signals typically hold an underlying lower-dimensional structure which is significant statistically~\cite{common, field}. Dictionary learning aims at obtaining the optimal set of subcomponents, termed dictionary and representations by maximizing the mutual information between a set of substructures and the original signals~\cite{Wang, 789, 258082}.   Watanabe~\cite{WATANABE1981381} believed that finding a sparse dictionary is equivalent to extracting the hidden sparse structure by minimizing entropy.  Many researchers have leveraged  DL for image denoising~\cite{zheng_deep_2021}, image super-resolution~\cite{castro2023exploring}, and so on~\cite{, tang2023explainable, 7903603}. Additionally, DL is also applied in bioinformatics, such as medical image analysis~\cite{survey2}. A dictionary learning based deep-learning model is used for DNA motif detection~\cite{dicmotif}, but lacks of interpretability because of the black-box model. 

\subsection{Applications and Evaluation of DNA Representations}

High-quality DNA representation schemes not only contribute to decoding the biochemical information embedded within DNA sequences but also facilitate various DNA-related tasks such as DNA classification~\cite{LI2006135, ATT1, noauthor_frontiers_nodate}, DNA clustering~\cite{RANDIC2003202, RANDIC20031, BAI2007282}, and motif detection~\cite{Chu2022.11.06.515322}, among others. By accurately capturing the underlying structure and meaning of DNA sequences, the representation scheme can provide valuable insights into biological processes and facilitate various genomic analyses. For example, researchers designed an efficient coding technique inspired by Huffman coding to compute DNA sequence similarity~\cite{JIN2016325}. Therefore, the performance of downstream applications is one of the common evaluation metrics for the effectiveness of representation schemes.

\subsection{Explainability in DNA Representation Schemes}

Biologists have concluded several criteria to judge whether a representation scheme is well-designed, including accuracy, robustness, succinctness, and adaptability~\cite{JIN2017342}. Explainability in DNA representation learning refers to interpreting the biochemical semantic meaning under captured features. Previous DNA representation schemes primarily relied on explainable artificial intelligence (XAI) to enhance explainability, such as post-hoc techniques~\cite{SANTORSOLA20231}. For instance, Li Xue et al.~\cite{doi:10.1021/acs.jcim.8b00368} utilized convolutional kernel visualization to identify RNA motifs. Besides, \textbf{AttCRISPR}~\cite{PMID:34903170} incorporates attention modules into the model to indicate the model's decisions at the global and local levels. Additionally, some metrics have also been designed to measure the contribution of each element in the representation towards the result of specific applications~\cite{10.1093/nar/gku1019}. However, these post-hoc techniques are challenging for biologists to validate explanations and comprehend insights from the representations.

Hence, our scheme aims to design a DL-based DNA representation scheme to represent DNA sequences as the concatenation of extracted semantic dymers, which embody the explainable representations and facilitate biologists' understanding by utilizing captured semantic dymers.

\section{Preliminary}



As highlighted by Badri et al.~\cite{6909690}, integrating sparsity constraints significantly improves the robustness of representations, especially in handling outliers and noise inherent in the data. Our scheme utilizes a dictionary learning framework to obtain sparse and explainable representations for DNA sequences. Therefore, this section provides the necessary preliminary knowledge about sparse representation and dictionary learning.

\subsection{Sparse Representation Learning}

Sparse representation learning serves as a potent technique for representing signals of interest by harnessing sparse vectors derived from a well-designed dictionary~\cite{candes2006robust, donoho2006compressed}. In other words, it entails identifying and reconstructing a signal $y \in \mathbb{R}^{d_y}$ characterized by a sparse representation $x \in \mathbb{R}^{d_x}$ within a high-dimensional space $\Phi \in \mathbb{R}^{d_y*d_x}: \Phi = [\phi_1, \dots, \phi_{d_x}]$. Essentially, the problem can be conceptualized as an optimization problem:

\begin{equation}
\begin{aligned}
    & \mathop{\argmin}\limits_{x \in \mathbb{R}^{d_x}} \psi(x) \\
    & s.t.\ \Phi x = y
\end{aligned}
\end{equation}
Here, $\psi(x)$ represents a sparsity constraint on $x$.




It is important to note that sparse representation learning and sparse recovery, while related, address different problems. As clarified in recent literature~\cite{arora_simple_2015}, sparse recovery focuses on recovering a sparse vector from under-determined linear measurements, while sparse representation learning seeks to find a sparse vector based on a dictionary to represent a non-sparse vector. Our work primarily solves the sparse representation learning problem.

\subsection{Dictionary Learning}

A typical sparse representation learning framework is Dictionary Learning (DL), which aims at learning a well-constructed dictionary $D \in \mathbb{R}^{d_y*d_x}$ to map each input data $y \in \mathbb{R}^{d_y}$ as a linear combination of dictionary elements. A high-quality dictionary should consist of $n$ elements $D = [d_1,\dots,d_{n}]$ that effectively approximate the original data and transform each input data into a sparse representation $x \in \mathbb{R}^{d_x}$~\cite{7102696}, which could be formulated as the following optimization problem:

\begin{equation}
\begin{aligned}
    & \mathop{\argmin}\limits_{D\in\mathcal{C}, x \in \mathbb{R}^{d_x}} \left\|y-Dx\right\|_2^2 + \lambda \psi(x), \\
    where \ \mathcal{C} \equiv & \left\{D \in \mathbf{R}^{d_y*d_x}: \left\|d_i\right\|_2 \leq 1,\ \forall i=1,...,n \right\}
\end{aligned}
\end{equation}

Various algorithms have been devised to optimize the original multivariate optimization problem alternatively~\cite{7102696}. Among the alternating optimization techniques, Coordinate Gradient Descent (CGD)~\cite{coor, coor2, cgd, cgd2} stands out for its simplicity and ability to handle large-scale problems efficiently. 

However, these existing methods exhibit two main limitations. First, the representations they produce are typically 1-dimensional, which leads to a loss of positional information. Second, the dictionaries they rely on tend to be redundant, potentially reducing the effectiveness of the representations.




\section{Explainable DNA Representation Structure}

This section intends to demonstrate how our model, \textbf{Dy-mer}, constructs the explainable representation structure by representing DNA sequences as a concatenation of several dymers. Firstly, some imperative concepts should be clarified.

A DNA sequence is usually a string of $l$ permuted nucleotides $b_i \in \{A, T, G, C\}, 0 \leq i \leq l$. A substring of length $k$ within a DNA $d$ is denoted as a dymer candidate $\ b_{1}b_{2}\dots b_{k} \ where\ k \in \mathbb{Z}^{+}$, the abbreviation for a K-mer of dynamic length. We could use one-hot code to represent DNA and dymer with matrices $d \equiv \left[\delta(b_{1})\ \delta(b_{2})\ \dots\ \delta(b_{l})\right]_{4\times l}$ and $\phi \equiv \left[\delta(b_{1})\ \delta(b_{2})\ \dots\ \delta(b_{k})\right]_{4\times k}$. Here, $\delta(b_{i})$ denotes the one-hot mapping. For example, \textit{AATTCGAT} could be written as a matrix $d$, where each row represents a type of nucleotide and each column represents a position. Therefore, the matrix of a 3-mer in \textit{AATTCGAT}, such as \textit{AAT} is
\begin{equation}
\begin{aligned}
    \textit{\textbf{AAT}}TCGAT \qquad \qquad \Rightarrow & \ \ \quad \quad AAT \\
    \begin{bmatrix}
    1 & 1 & 0 & 0 & 0 & 0 & 1 & 0 \\ 
    0 & 0 & 1 & 1 & 0 & 0 & 0 & 1 \\
    0 & 0 & 0 & 0 & 0 & 1 & 0 & 0 \\
    0 & 0 & 0 & 0 & 1 & 0 & 0 & 0 \\
    \end{bmatrix} \quad \Rightarrow & \quad
    \begin{bmatrix}
    1 & 1 & 0 \\ 
    0 & 0 & 1 \\
    0 & 0 & 0 \\
    0 & 0 & 0 \\
    \end{bmatrix}
\end{aligned}
\end{equation}

\begin{figure*}[htbp]
\centering
\includegraphics[width=1\textwidth,height=0.4\textwidth]{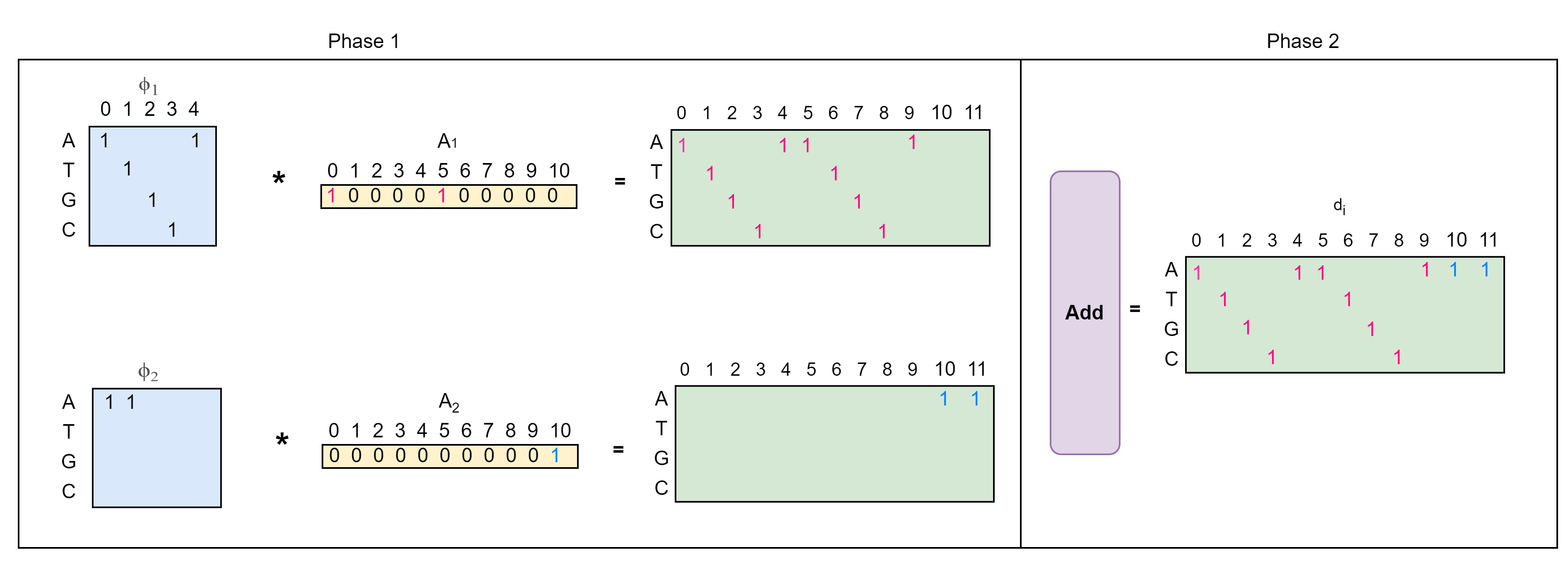}
\caption*{Reconstruction follows a 2-phase process. Firstly, 1D convolution operations are conducted on each K-mer with its corresponding row of the assignment matrix. In this example, the red and blue colors identify two different K-mers. Then, add up all the results and get the reconstructed DNA sequence matrix.}
\caption{An example for DNA reconstruction}
\label{4.2}
\end{figure*}

Studies indicate that biochemical properties and functions are often associated with similar dymers, a.k.a. motifs, which are typically represented with a position weight matrix (PWM) $\textbf{P} \in [0, 1]^{4\times l}$. A position weight matrix illustrates the positional distribution of nucleotides within a motif. Similarly, we can expand our dymer matrices into real-numbered matrices. Each element $p_{ij} \in [0,1]$ means the likelihood that $i_{th}$ nucleotide appears at the $j_{th}$ position, which offers greater generalizability.

Our model intends to represent DNA sequence $d_i$ as a concatenation of several dymers, $\hat{d}_i = \phi_{1}\phi_{2}\dots \phi_{s}, where\ \phi_j \in \Phi, 0 \leq j \leq s$. The assignment matrix is defined to record the assignment of dymers to their respective positions. 

\begin{myDef}[Assignment Matrix]
    Given a DNA sequence $d$ and a basis $\Phi$, an assignment matrix $\textbf{A} \in \{0, 1\}^{n\times l}$ is defined as an $n \times l$ matrix. Each row represents a dymer $\phi$ from $\Phi$, while each column represents a position in $d$. At $i_{th}$ row and $j_{th}$ column of $\textbf{A}$, $a_{ij}$ is a binary variable meaning whether the $i_{th}$ dymer $\phi_i$ starts at position $j$ in the given DNA.
\end{myDef}

Similarly, if we turn the assignment matrix into a real matrix, each entry indicates the likelihood of selecting the corresponding dymer at a specific position. 

However, there are various combinations of dymers to reconstruct the original DNA sequence. We define the collection of all possible dymers for DNA $d$ as the dymer spectrum $\hat{\Phi}$. A subset of $\hat{\Phi}$ containing $n$ distinct dymers for DNA representation is termed a dymer dictionary $D$. Given a $D$ and $\textbf{A}$, we seek to define an operator that reconstructs the original DNA sequence as the concatenation of dymers. Since the dymer matrix encodes both positional and sequential information, a convolution operator is a natural and effective choice for modeling the concatenation process. 



Given a DNA sequence $d$ with $l$ nucleotides and a well-constructed dictionary $D$ of $n$ dymers, the DNA sequence matrix $d$ has dimensions $4 \times l$, and each dymer matrix $\phi \in D$ has dimensions $n \times L$, where $L$ means the length of the longest dymer. And the assignment matrix $\textbf{A}^{d}$ for DNA $d$ is obtained. With these elements in place, the DNA sequence can be reconstructed using a two-step operation, a 1D convolution operation followed by a sum-up operation. See Figure \ref{4.2} for a specific example.
\begin{equation}
 \begin{aligned}
     \hat{d} & = \textbf{A}^{d} \odot D \\
     & \triangleq \left[ \sum_{\phi\in D} \phi_{b} \ast \textbf{A}^{d}_{\phi} \right]_{b=1}^4
 \end{aligned}
\end{equation}
Here, $\phi$ represents a dymer, and $\phi_{b}$ is the $b_{th}$ row of the matrix $\phi$, on behalf of the $b_{th}$ nucleotide. $\ast$ is the 1D convolution operator. $\left[\bullet\right]^B_b$ is the stacking operation that stacks the inner vectors $\bullet_b, \bullet_{b+1}, \dots, \bullet_B$ along a new dimension. $\textbf{A}^{d}_{\phi}$ denotes the corresponding row to K-mer $\phi$ in $\textbf{A}^d$.  There are several variations of $\textbf{A}^d$. Considering $m$ assignment matrices of $n$ DNA $\{d_1, \dots, d_m\}$ based on a dictionary $D$, we can collect the row corresponding to K-mer $\phi \in D$ from each $\textbf{A}^d$ and stack them along a new dimension as $\textbf{A}_{\phi}$, which describes the assignment of K-mer $\phi$ towards each DNA.


If one assignment matrix could reconstruct the DNA sequence accurately, it records the correct composition and permutation of dymers within the DNA, which could serve as an explainable representation.

\section{Dymer Dictionary Learning for DNA Representations}

\subsection{Problem Formulation}

Given $m$ DNA sequences $\{d_1, d_2, \dots, d_m\}$ and a well-constructed basis $\Phi$ consisting of $n$ semantic K-mers.  According to the dictionary learning framework, the optimal assignment matrices $\textbf{A}^{d_i} \in \{\textbf{A}^{d_1}, \textbf{A}^{d_2}, \dots, \textbf{A}^{d_m}\}$ could be obtained by solving the following optimization problem,



\begin{equation}
\begin{aligned}
    \mathop{\argmin}\limits_{\phi \in \textbf{D}, \phi \in [0,1]^{4\times L} \textbf{A}^{d_i} \in A, \textbf{A}^{d_i} \in \{0,1\}^{n\times l}} & \sum_{i=1}^m \left\|d_i-\hat{d}_i\right\|_2^2 + \lambda_A \sum_{i=0}^{m} \Psi(A^{d_i}), \\
\end{aligned}
\label{eq:4.41}
\end{equation}

Here, $\Psi$ represents a sparse penalty on the assignment matrix, and $\hat{d}$ denotes the reconstructed DNA sequences.

The objective function incorporates two components: the sparsity constraints and the accuracy constraint. The accuracy constraint ensures the precise reconstruction of each DNA sequence. While multiple assignment matrices may satisfy this constraint, many might select insignificant or meaningless dymers due to the inherent complexity and noise of DNA data, thus compromising the robustness of representations. Therefore, sparsity constraints are defined to achieve the following objectives:

\begin{itemize}
    \item The dictionary should be compact and low-rank by selecting frequent and significant dymers while eliminating noise.
    \item Each DNA representation (the assignment matrix) should be sparse and succinct.
\end{itemize}

\subsection{Constraint Definitions}

\begin{figure*}[htbp]
\centering
\begin{minipage}[h!]{1\textwidth}
\centering
\includegraphics[width=1\textwidth,height=0.38\textwidth]{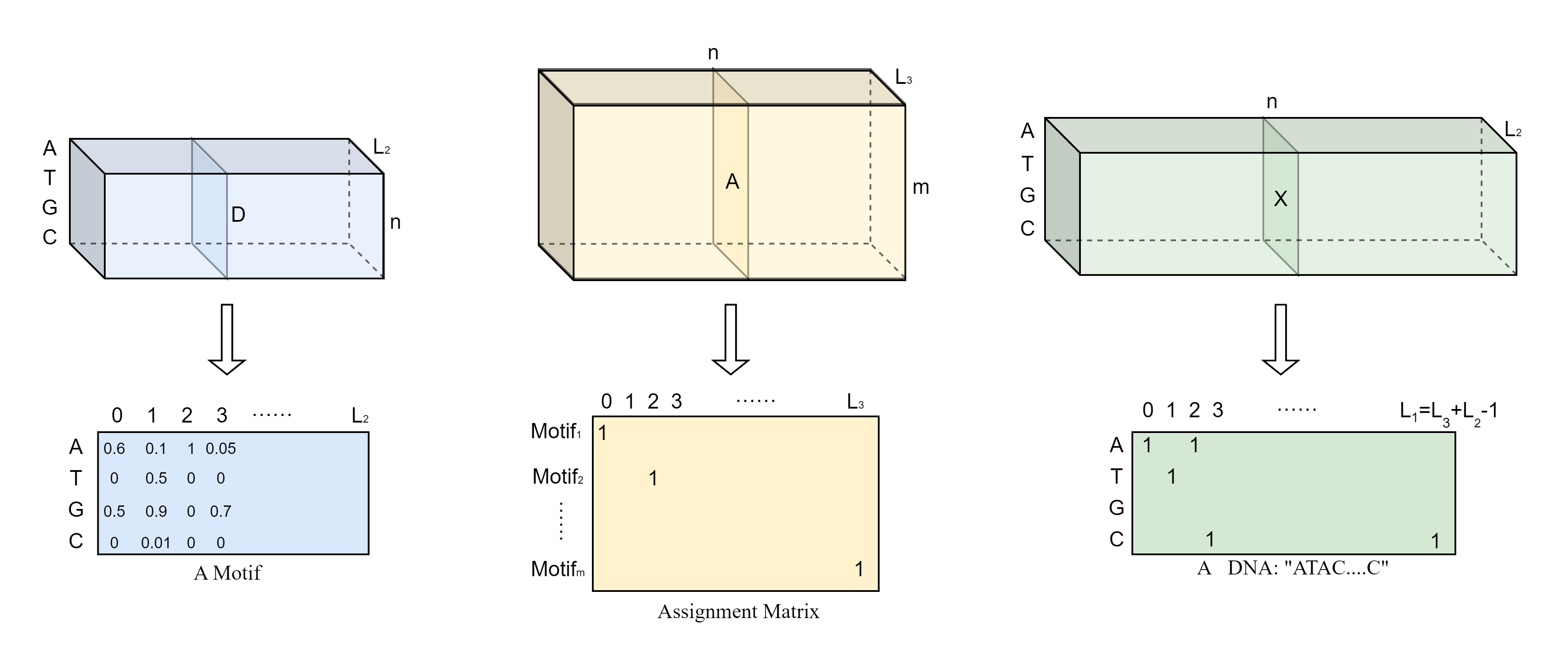}
\caption{An illustration of tensors in sparse representation based DNA representation scheme}
\label{5.1}
\end{minipage}
\end{figure*}

To enhance the computational efficiency for batch data processing, we formulate the constraints in tensor format. 

All notations are illustrated in Figure \ref{5.1}. The rightmost green 3-dimensional tensor $\mathcal{X} \in \{0,1\}^{m\times 4\times L_1}$ represents $m$ DNA sequence matrices $\{d_1, d_2, \dots, d_m\}$, where 4 represents the four different nucleotides (A, T, G, C) and $L_1$ represents the index number in DNA sequences. Each layer along the first dimension of this tensor corresponds to the DNA sequence $d_i$ with padding 0. In addition, $\hat{\mathcal{X}} \in \{0,1\}^{m\times 4\times L_1}$ is a tensor shaped as $\mathcal{X}$ to store the reconstructed DNA sequences. 

The leftmost blue 3-dimensional tensor $\textbf{D} \in \{0,1\}^{n\times 4\times L_2}$ represents the dictionary, where $L_2$ denotes the maximum length among all dymers. Each layer along the first dimension of these tensors corresponds to the dymer matrix $\phi$ with padding 0. 

The 3-dimensional tensor $\mathcal{A} \in \{0,1\}^{n\times m\times L_3}$ in the middle represents the stacked assignment matrices, where $m$ and $n$ represent the number of DNA sequences and different dymers and $L_3$ represents the maximum position where a dymer can be placed of every DNA sequence. Each layer along the first dimension of this tensor corresponds to the matrix $ \textbf{A}_{\phi}$ with padding 0. Each layer along the second dimension of this tensor corresponds to the assignment matrix $\textbf{A}^{d_i}$ with padding 0. 

According to the rules of convolution, $L_1 = L_2 + L_3 - 1$.

\subsubsection{Accuracy Constraint}


Initially, it's crucial to represent DNA sequences accurately. Therefore, the disparity between the reconstructed DNA sequence matrix and the original DNA sequence matrix should be minimized. We denote the euclidean distance between $\mathcal{X}$ and $\hat{\mathcal{X}}$, $dist(\mathcal{X}, \hat{\mathcal{X}})$, is defined as the Reconstruction Loss $\mathcal{L}_{RC}$.

\begin{equation}
    \begin{aligned}
        \mathcal{L}_{RC} & = dist(\mathcal{X}, \hat{\mathcal{X}}) \\ 
        & = \sum_{i=1}^{m} \left\|d_i - \hat{d_i}\right\|^2_2 
        = \sum_{i=1}^{m} \left\|d_i - \textbf{A}^{d_i} \odot \textbf{D} \right\|^2_2 \\
        & = \sum_{i=1}^{m}\left\|d_i -\left[\sum_{\phi\in \Phi} \phi_{b} \ast \textbf{A}_{\phi}^{d_i} \right]_{b=1}^4\right\|_2^2\
    \end{aligned}
\label{eq:4.5}
\end{equation}

Here, $\|\bullet \|^2_2$ denotes the squared second norm.

The tensor-based formulation of $L_{RC}$ is 
\begin{equation}
    \begin{aligned}
        \mathcal{L}_{RC} & = \left\|\mathcal{X} - \hat{\mathcal{X}}\right\|^2_2 = \left\|\mathcal{X} - \mathcal{A} \odot \textbf{D} \right\|^2_2 \\
         whe&re \ \hat{\mathcal{X}}_{ijl} = \sum_{k} \sum_{s} \mathcal{A}_{kis} \bullet \textbf{D}_{kj(l-s)}
    \end{aligned}
\label{eq:4.9}
\end{equation}


\subsubsection{Sparsity Constraints}

Sparsity constraints can handle outliers and noise and capture frequent dymers to enhance the robustness of representations. Based on the mentioned objectives, two sparsity constraints can be designed separately. 

To ensure the selection of significant dymers and exclude noisy or redundant ones, a sparsity constraint termed Dictionary Complexity ($L_{DC}$) is introduced. This constraint quantifies the count of significant dymers in the dictionary $\textbf{D}$ to penalize noise or redundant dymers, thus contributing to a more compact and lower-rank dictionary. 

The significance of a dymer is usually assessed by its frequency in DNA sequences. For each DNA $d_i$ and a dymer $\phi$, the maximum value of the vector $\textbf{A}^{d_i}_{\phi}$ serves as a binary variable indicating whether $\phi$ is utilized in $d_i$. Therefore, the maximum value of $\textbf{A}_{\phi}$ serves as an indicator of whether the dymer $\phi$ is utilized among the DNA dataset $\textbf{d}$.

Given a dymer dictionary $\textbf{D}$, the Dictionary Complexity is,

\begin{equation}
    \begin{aligned}
        \mathcal{L}_{DC} = \left|\textbf{D}\right| = \sum_{\phi \in \textbf{D}} \max  \textbf{A}_{\phi}
    \end{aligned}
\label{eq:4.6}
\end{equation}

However, the scheme may prioritize short and repetitive dymers to reduce the Dictionary Complexity. For example, a dictionary comprising only \textit{A, T, G, C} as dymers greatly decreases the Dictionary Complexity but fails to extract longer dymers. Therefore, another sparsity constraint aims at utilizing as few dymers as possible for each DNA, which ensures longer and more representative dymers. 

To achieve this goal, we define the Average Representation Loss $\mathcal{L}_{RE}$, which quantifies the average number of dymers to reconstruct DNA sequences. The assignment matrix $\textbf{A}^{d}$ records whether a dymer contributes to DNA $d$ reconstruction. Summing all the elements up yields the total number of dymers used. Hence, given $m$ assignment matrices for each DNA sequence ${\textbf{A}^{d_1}, \textbf{A}^{d_2}, \dots, \textbf{A}^{d_m}}$, we can then compute the Average Representation Loss by averaging the $L1$ norms of these matrices.

\begin{equation}
    \mathcal{L}_{RE} = \frac{\sum_{i=1}^{m} \|\textbf{A}^{d_i}\|_1}{m}
\label{eq:4.7}
\end{equation}

Here, $\|\bullet\|_1$ denotes the $L1$ norm of a matrix.

Using tensor notation, the Dictionary Complexity cost $\mathcal{L}_{DC}$ and the average representation cost loss $\mathcal{L}_{RE}$ are

\begin{equation}
    \begin{aligned}
        \mathcal{L}_{BC} & = \sum^{m}_{j=1} \max \mathcal{A}_{\phi_{j}::} \\ 
        & = \min ((\mathcal{A} \times_{1} \textbf{1}_{L_2}) \ \times_{1} \theta, \textbf{1}_{m}) \times_{1} \textbf{1}_m \\
       where \  \textbf{1}_m & = \begin{vmatrix} 1 & 1 & ...& 1 & 1 \end{vmatrix}^{T}_{m} \\
       \textbf{E}_n & = \begin{vmatrix} 
                    1 & 0 & ...& 0 & 0 \\
                    0 & 1 & ...& 0 & 0 \\
                    ... & ... & ...& ... & ... \\
                    0 & 0 & ... & 1 & 0 \\
                    0 & 0 & ...& 0 & 1
                \end{vmatrix}_{n\times n} \\
        \theta & = \textbf{E}_{n} \otimes \textbf{1}_{m}
    \end{aligned}
\label{eq:4.10}
\end{equation}

$\otimes$ denotes the Kronecker product and $\times_{1}$ denotes the left multiplication between a matrix and a $mode$-1 Matricization tensor.

\begin{equation}
    \mathcal{L}_{RE} = \frac{\sum_{i=1}^{m} \|\mathcal{A}_{:d_{i}:}\|_1}{m} = \frac{\|\mathcal{A}\|_1}{m}
\label{eq:4.11}
\end{equation}

Finally, given DNA sequences $\mathcal{X}$, and a dictionary $D$, the objective function could be written as follows. 

    
\begin{small}
\begin{equation}
    \begin{aligned}
        \mathcal{A}^* & = \mathop{\argmin} \ (\lambda_{RC}\ \mathcal{L}_{RC} + \lambda_{DC}\ \mathcal{L}_{DC} + \lambda_{RE}\ \mathcal{L}_{RE}) \\
        & = \mathop{\argmin}\limits_{\mathcal{A} \in \{0,1\}^{n\times m\times L_3}} \lambda_{RC} \sum_{i=1}^{n}\left\|\mathcal{X}_{:d_{i}:} -\left[ \sum_{j=1}^n \textbf{D}_{\phi_{j}b:} \ast \mathcal{A}_{\phi_{j} d_{i}:} \right]_{b=1}^4\right\|_2^2 \\
        & +\lambda_{DC} \sum_{j=1}^n 
       \mathop{\max} \mathcal{A}_{\phi_{j}::} + \lambda_{RE} \frac{\sum_{i=1}^{m} \|\mathcal{A}_{:d_{i}:}\|_1}{m} \\
        & = \mathop{\argmin}\limits_{\mathcal{A} \in \{0,1\}^{n\times m\times L_3}} \lambda_{RC} \left\|\mathcal{X} - \mathcal{A} \odot \textbf{D}\right\|_2^2 \\
        & +\lambda_{DC} \min ((\mathcal{A} \times_{1} \textbf{1}_{L_2}) \ \times_{1} \theta, \textbf{1}_{m}) \times_{1} \textbf{1}_m + \lambda_{RE} \frac{\|\mathcal{A}\|_1}{m}
    \end{aligned}
\label{eq:4.12}
\end{equation}
\end{small}

\begin{algorithm}[h!]
\caption{Dy-mer: Dictionary Learning for DNA Representation}
\label{alg:dymer_dl}
\begin{algorithmic}[1]
\renewcommand{\algorithmicrequire}{\textbf{Input:}}
\renewcommand{\algorithmicensure}{\textbf{Output:}}

\REQUIRE
  DNA sequence tensor $\mathcal{X} \in \{0,1\}^{m \times 4 \times L_1}$; \\
  Hyperparameters $\lambda_{RC}, \lambda_{DC}, \lambda_{RE}$; \\
  Number of epochs $T_{max}$; number of dymers $n$; maximum dymer length $L_2$.
\ENSURE
  Optimal dymer dictionary $\mathcal{D}^* \in [0,1]^{n \times 4 \times L_2}$; \\
  Optimal DNA representations $\mathcal{A}^* \in [0,1]^{n \times m \times L_3}$.

\STATE \textbf{Initialize:} Randomly initialize $\mathcal{D}_0 \in [0,1]^{n \times 4 \times L_2}$ and $\mathcal{A}_0 \in [0,1]^{n \times m \times L_3}$, where $L_3 = L_1 - L_2 + 1$.

\FOR{$t = 1$ \TO $T_{max}$}
    \STATE \textit{// 1. Sparse Representation Update}
    \STATE Fix $\mathcal{D}_{t-1}$, and update $\mathcal{A}_t$ by minimizing:
    $$
        \mathcal{L}_{\mathcal{A}} \leftarrow \lambda_{RC} ||\mathcal{X} - \mathcal{A} \odot \mathcal{D}_{t-1}||_2^2 + \lambda_{DC} \mathcal{L}_{DC}(\mathcal{A}) + \lambda_{RE} \frac{||\mathcal{A}||_1}{m}
    $$

    \STATE \textit{// 2. Dictionary Update}
    \STATE Fix $\mathcal{A}_t$, and update $\mathcal{D}_t$ by minimizing:
    $$
        \mathcal{L}_{\mathcal{D}} \leftarrow ||\mathcal{X} - \mathcal{A}_{t} \odot \mathcal{D}||_2^2
    $$
\ENDFOR

\STATE \textbf{return} $\mathcal{D}_{T_{max}}$, $\mathcal{A}_{T_{max}}$.
\end{algorithmic}
\end{algorithm}


\subsection{Optimization Algorithm}
Then, optimizing the joint optimization  \eqref{eq:4.12}  simultaneously is nearly infeasible. Consequently, we propose an alternating optimization approach to tackle this multivariate optimization challenge, including two primary phases: sparse representation learning and dictionary update. 



At each time step $t$, we begin by fixing the dictionary $\mathcal{D}_{t-1}$ from the previous iteration and then determine the optimal assignment tensor $\mathcal{A}_{t}$:

\begin{small}
\begin{equation}
    \begin{aligned}
        \mathcal{A}_{t} & = \mathop{\argmin}\limits_{\mathcal{A}_{t} \in [0,1]^{n\times m\times L_3}} \lambda_{RC} \left\|\mathcal{X} - \mathcal{A}_{t-1} \odot \mathcal{D}_{t-1}\right\|_2^2 \\
        & +\lambda_{DC} \min ((\mathcal{A}_{t-1} \times_{1} \textbf{1}_{L_2}) \ \times_{1} \theta, \textbf{1}_{m}) \times_{1} \textbf{1}_m + \lambda_{RE} \frac{\|\mathcal{A}_{t-1}\|_F}{m}
    \end{aligned}
\label{eq:5.11}
\end{equation}
\end{small}

Speaking of convexity, the convolution operation and $L_1$ norm are essentially linear operations among elements of operands, which could be simplified into a quadratic function. The objective function could be viewed as a quadratic term along with a maximizing term on the elements of $\mathcal{A}$, both of which are convex. Given the linearity of convexity and the convex input variable domain, the objective function is convex. After solving these problems, we update the assignment matrix $\mathcal{A}_t$ once.


Then, the dictionary $\mathcal{D}_{t}$ is updated by fixing the assignment tensor  $\mathcal{A}_{t}$.

\begin{small}
\begin{equation}
    \begin{aligned}
        \mathcal{D}_{t} & = \mathop{\argmin}\limits_{\mathcal{D}_{t} \in [0,1]^{n\times m\times L_3}} \lambda_{RC} \left\|\mathcal{X} - \mathcal{A}_{t} \odot \mathcal{D}_{t-1}\right\|_2^2 \\
    \end{aligned}
\label{eq:5.12}
\end{equation}
\end{small}

Similarly, the optimization problem can be simplified to a quadratic optimization, which is convex.

This iterative process is repeated until the loss converges or both the dictionary and the assignment tensor become stable. Finally, we get the optimal dictionary $\mathcal{D}^{*}$ and assignment tensor $\mathcal{A}^{*}$. However, it is not necessary to solve the optimal solution in each iteration, because the time complexity will greatly increase as the data size grows. For simplicity, we use the coordinate gradient descent, mentioned in section 2.5, as a substitution. 


\subsection{Obtain Frequent dymer Dictionary by Thresholding}

After obtaining the optimal dictionary $\mathcal{D}^*$, frequent dymers could be selected according to the assignment tensor $\mathcal{A}^*$.

Firstly, we compute the average frequency $\mathcal{L}_{\phi_j}$ of each dymer $\phi_j$ and construct a vector of frequency for all dymers $\mathcal{L} = \left[\mathcal{L}_{\phi_1}, \mathcal{L}_{\phi_2}, ..., \mathcal{L}_{\phi_n}\right]$.

\begin{small}
    \begin{equation}
        \begin{aligned}
            \mathcal{L}_{\phi_j} = \frac{\sum_{j=1}^{n} (\mathcal{A}^{*}_{\phi_j::})}{m}
        \end{aligned}
        \label{eq:5.13}
    \end{equation}
\end{small}

Then, we set a threshold and use the indicator function $\Delta_{\epsilon}$ to transfer the frequency vector into a binary indicator vector $\mathcal{I} = \Delta_{\epsilon}(\mathcal{L})$. Finally, frequent dymers are collected as an $\epsilon$-frequent dymer dictionary $\epsilon$-$\mathcal{D}^*$.

\begin{small}
    \begin{equation}
        \begin{aligned}
             \mathcal{D}^* & = \mathcal{D} \times_{1} (\mathcal{I} \otimes \textbf{I}_{n})
        \end{aligned}
    \label{eq:5.14}
    \end{equation}
\end{small}

Here, $\textbf{I}_{n}$ represents a vector of length $n$ with all elements being 1.


Since we obtained $\epsilon$-frequent dymer dictionary $\epsilon$-$\mathcal{D}^*$, we can represent unseen DNA sequences with these frequent dymers. Firstly, we get the tensor of new DNA sequences, denoted as $\mathcal{X}_{new}$.  Then, the representations are obtained by solving the following optimization problem and deriving the optimal assignment matrix for each DNA sequence $d_i$.

\begin{small}
    \begin{equation}
        \begin{aligned}
            \mathcal{A}^* & = \mathop{\argmin}_{\mathcal{A}} \ (\lambda_{RC}\ \mathcal{L}_{RC} + \lambda_{DC}\ \mathcal{L}_{DC} + \lambda_{RE}\ \mathcal{L}_{RE}) \\
            s.t. 
            & \quad \mathcal{D} = \mathcal{D}^*, \ \mathcal{X} = \mathcal{X}_{new}
        \end{aligned}
    \label{eq:5.15}
    \end{equation}
\end{small}



\section{Experiments}

In this section, we evaluate the effectiveness and explainability of our representation scheme through a critical DNA-related task: DNA promoter classification. Subsequently, we can capitalize on the explainability of our representation scheme for various downstream applications, including DNA clustering and motif detection. 


\subsection{DNA promoter classification}

\subsubsection{Experiment Background}

DNA promoters play a crucial role in gene expression regulation and transcription, making accurate identification of these segments essential for understanding various biological processes.


Researchers have identified several validated motifs that commonly appear in promoters, including CpG islands~\cite{cpg}, CCAAT box~\cite{caat}, TATA Box~\cite{tata}, GC box~\cite{gcbox} and transcriptional initiator~\cite{ini}. Promoters, especially strong ones, tend to be rich in GC regions compared to non-promoters. These motifs are typically represented using the IUPAC one-letter codes, where \textit{A}, \textit{T}, \textit{G}, \textit{C} denote their respective nucleotides. \textit{R} denotes purine(A/G), and \textit{Y} signifies Pyrimidines(T/C) and \textit{W} represents Weak interaction (A/T). An effective and explainable DNA representation scheme should be capable of selecting various significant dymers like these mentioned motifs as indicators. 



\begin{table*}[!h]
\centering
\caption{Comparison to existing methods on DNA promoter classification}
\setlength{\tabcolsep}{5mm}
\begin{tabular}{l*{4}{c}}
 \toprule
 Model & \textbf{SENS} & \textbf{SPEC} & \textbf{ACC} & \textbf{MCC} \\
 \hline
 \textbf{First Task} & & & & \\
 Ours & \textbf{96.90} & \textbf{99.34} & \textbf{98.10} & \textbf{96.25} \\
 Fasttext+CNN~\cite{noauthor_frontiers_nodate} & 82.76 & 88.05 & 85.41 & 70.90 \\
 IPSW(2L)-PseKNC~\cite{xiao_ipsw2l-pseknc_2019} & 81.32 & 84.89 & 83.13 & 66.30 \\
 iPromoter-2L~\cite{10.1093/bioinformatics/btx579} & 79.20 & 84.16 & 81.68 & 63.43 \\
 iPro54~\cite{10.1093/nar/gku1019} & 77.76 & 83.15 & 80.45 & 61.00 \\
 Stability~\cite{DEAVILAESILVA201422} & 76.61 & 79.48 & 78.04 & 56.16 \\
 vw Z-cuver~\cite{article} & 77.76 & 82.80 & 80.28 & 60.98 \\
 PCSF~\cite{LI2006135} & 78.92 & 70.70 & 74.81 & 49.80 \\
  DNABERT (Finetune All Parameters)~\cite{10.1093/bioinformatics/btab083} & 79.20 & 78.20 & 78.66 & 57.35 \\
  DNABERT (Finetune Last Layer) & 73.80 & 73.76 & 73.75 & 47.52 \\
  DNABERT-2 (Finetune All Parameters)~\cite{zhou2024dnabert} & 83.81 & 81.43 & 82.55 & 65.17 \\
  DNABERT-2 (Finetune Last Layer) & 75.30 & 71.87 & 73.45 & 47.07 \\
  GROVER (Finetune All Parameters)~\cite{sanabria2024dna} & 83.54 & 80.90 & 82.11 & 64.35 \\
  GROVER (Finetune Last Layer) & 80.39 & 73.78 & 76.70 & 53.83 \\
 \textbf{Second Task} & & & & \\
 Ours & \textbf{94.16} & \textbf{93.51} & \textbf{93.83} & \textbf{87.71} \\
 Fasttext+CNN~\cite{noauthor_frontiers_nodate} & 69.40 & 79.17 & 73.10 & 46.00 \\
 IPSW(2L)-PseKNC~\cite{xiao_ipsw2l-pseknc_2019} & 62.23 & 79.17 & 71.20 & 42.13 \\
  DNABERT (Finetune All Parameters)& 53.82 & 56.23 & 55.37 & 10.89 \\
  DNABERT (Finetune Last Layer) & 49.84 & 54.10 & 52.83 & 6.64 \\
  DNABERT-2 (Finetune All Parameters)& 58.47 & 55.26 & 55.38 & 11.35 \\
  DNABERT-2 (Finetune Last Layer) & 53.05 & 53.07 & 53.01 & 5.57 \\
  GROVER (Finetune All Parameters)& 48.96 & 55.46 & 54.19 & 17.67 \\
  GROVER (Finetune Last Layer) & 45.57 & 52.78 & 51.72 & 4.32 \\
 \bottomrule
\end{tabular}
\label{tab5.1}
\end{table*}

\begin{table*}[h]
\centering
\caption{The result of the 10-fold cross-validation}
\setlength{\tabcolsep}{6mm}
\begin{tabular}{l*{4}{c}}
  \toprule
   & \textbf{SENS} & \textbf{SPEC} & \textbf{ACC} & \textbf{MCC} \\
  \hline
  \textbf{First Task} & & & & \\
  Mean & 96.90 & 99.34 & 98.10 & 0.9625 \\
  Standard Deviation & 0.0151 & 0.0065 & 0.0069 & 0.0013 \\
  \textbf{Second Task} & & & & \\
  Mean & 94.16 & 93.51 & 93.83 & 0.8771 \\
  Standard Deviation & 0.0244 & 0.0263 & 0.0170 & 0.0338 \\
  \bottomrule
\end{tabular}
\label{tab5.2}
\end{table*}

\subsubsection{Dataset}

\textbf{Promoter/non-Promoter}. A high-quality DNA promoter dataset collected by Xiao et al.~\cite{xiao_ipsw2l-pseknc_2019}, sourced from RegulonDB~\cite{10.1093/nar/gkv1156}, a well-established database of the regulatory network of gene expression. This dataset comprises 6,764 DNA sequences, each 81 base pairs (bp) long. Half of these sequences are DNA promoter samples, which are specific regions that initiate the transcription process. The remaining sequences are non-promoters. Within the promoter samples, there are 1,591 strong promoters and 1,792 weak promoters.

The dataset was divided into training and testing sets following a ratio of 2:1. Each subset contains three classes of data in equal proportion. This division across classes ensures data balancing and mitigates bias during model training and evaluation. 

\subsubsection{Experiment Setup}

There are two primary tasks: firstly, classifying promoters from the remaining non-promoter sequences; secondly, categorizing the promoters into strong promoters and weak promoters.  

\begin{figure}[htbp]
\centering
\begin{minipage}[t]{0.5\textwidth}
\centering
\includegraphics[width=1\textwidth,height=0.27\textwidth]{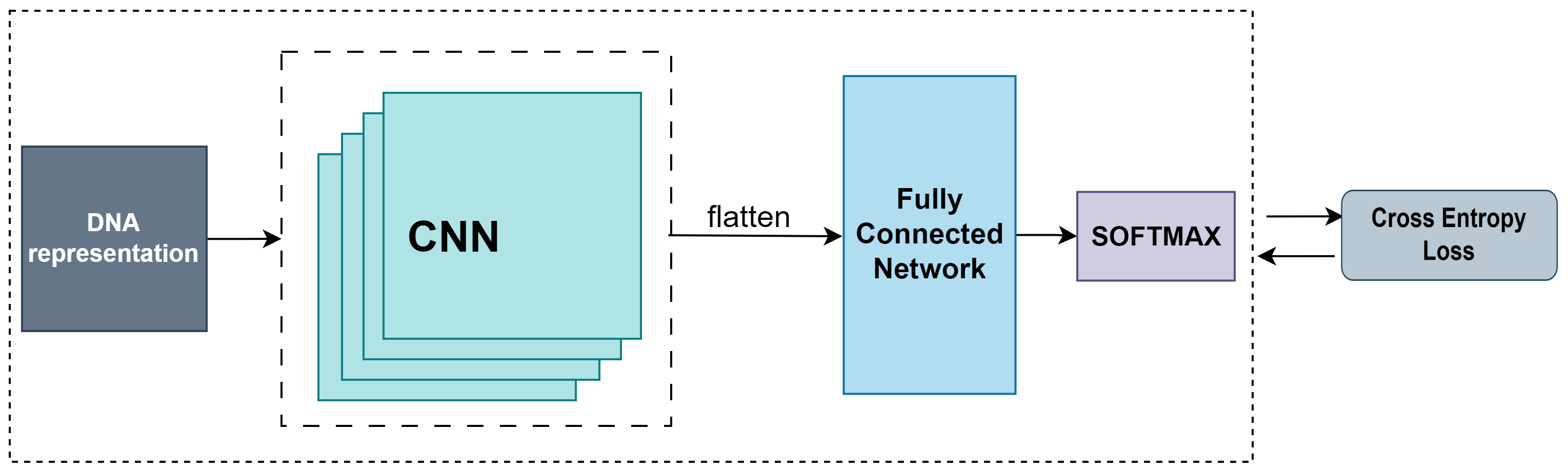}
\caption{An illustration of DNA promoter classification model}
\label{5.5}
\end{minipage}
\end{figure}

Therefore, we input the representations into a Convolutional Neural Network (CNN)-based model for the classification, as depicted in Figure \ref{5.5}. Based on the results on the validation set, we set the number of dymer in the dictionary as 1,400 and the Lagrange multipliers as $\lambda_{RC}:\lambda_{BC}:\lambda_{RE} = 1:1:1$.


We use 4 indicators to evaluate the model performance.

\begin{myDef}
    \begin{enumerate}[(i)]
    \item \textbf{Sensitivity(Sens)}:
    $1-\frac{N^+_-}{N^+}, 0 \leq Sen \leq 1$
    
    \item \textbf{Specificity (Spec)}:
    $1-\frac{N^-_+}{N^-}, 0 \leq Spec \leq 1$
    
    \item \textbf{Accuracy (Acc)}:
    $1-\frac{N^-_++N^+_-}{N^-+N^+}, 0 \leq Acc \leq 1$

    \item \textbf{Matthews Correlation Coefficient (MCC)}:

    $\frac{1-(\frac{N^-_++N^+_-}{N^-+N^+})}{\sqrt{(1+\frac{N^-_+-N^+_-}{N^+})(1+\frac{N^+_--N^-_+}{N^-})}}, -1\leq MCC \leq 1$
    \end{enumerate}
\end{myDef}

\begin{figure*}[htbp]
\centering
\end{figure*}

\subsubsection{Numerical Result}


Multiple experiments were conducted to benchmark the proposed representation scheme against existing methods. As reported in Table \ref{tab5.1}, the proposed scheme consistently surpasses all prior state-of-the-art approaches across all evaluated metrics on the benchmark dataset. Moreover, the 10-fold cross-validation results summarized in Table \ref{tab5.2} demonstrate the model's stability and robustness.

Two primary reasons are proposed to account for the observed performance advantage. First, the scheme effectively extracts salient dymers, serving as meaningful indicators for representing DNA sequences. We analyzed the top-50 frequent dymers drawn from the dymer dictionary for three categories: strong promoters, weak promoters, and non-promoters. Key findings are as follows:

\begin{enumerate}[(i)]
    \item In promoter sequences (both weak and strong), there is a pronounced enrichment of  \textbf{GC-rich regions (GC Box)}, especially in strong promoters. For example, K-mers like \textit{GCGCGC}, \textit{CGCGCG} are $\mathbf{3}$ times more prevalent in promoters compared to non-promoters. \textit{CGCGCGC} is even $\mathbf{15}$ times more frequent in promoters.  Additionally, K-mers such as \textit{GGCCCG}, \textit{GCCGCA}, \textit{GGGCGG} exhibit a higher likelihood in strong promoters compared to weak promoters.
    
    \item In promoters, the \textbf{TATA box} region is characterized by frequent occurrences of dymers like \textit{TATAAAA}, \textit{TATAAAT}, \textit{TATAAA} and \textit{TATAAT}. Similarly, the \textbf{CCAAT box} region exhibits enrichment with dymers such as \textit{GCCAAT}, \textit{CAATCT}.
    
    \item Furthermore, dymers resembling the \textbf{initiator elements} motif \textbf{YYANWYY}, such as \textit{TCACTCC}, \textit{TTAAATT}, \textit{TTATTTT}, \textit{CCATTTT} and \textit{CTATTTT}, appear at least $\mathbf{3}$ times more frequently in promoters than in non-promoters.
\end{enumerate}

Secondly, when compared with large-scale models that encompass significantly more parameters and are typically pre-trained on substantially larger corpora, our approach exhibits superior generalization. This superior performance is likely attributable to reduced overfitting on our relatively small dataset (6,766 sequences).

\subsubsection{Ablation Experiments}

To assess the necessity of the two sparsity constraints, we conducted ablation experiments isolating each constraint and evaluated their impact on the classification accuracy for the first task. The results are reported in Table \ref{tab5.5}. The table shows that both sparsity constraints contribute to improvements in classification accuracy, confirming their complementary role in enhancing the quality of the representation. 

\begin{table}[!h]
\centering
\caption{Results for ablation experiments}
\setlength{\tabcolsep}{5mm}
\begin{tabular}{cc}
  \toprule
   & \textbf{ACC} \\
  \hline
  \textbf{Original scheme} & \textbf{96.0\%} \\
  Scheme without basis complexity loss & 94.0\% \\
  Scheme without representation cost loss & 95.0\% \\
  \bottomrule
\end{tabular}
\label{tab5.5}
\end{table}

\subsubsection{Experimental Evaluation: Overcomplete vs. Learnable Dictionaries
}
Sometimes, in practice, a pre-defined dymer dictionary may already contain all the recognized significant dymers. Sparse representation can then be derived via a sparse representation learning-based scheme on this predetermined and overcomplete dictionary. However, the learnable dictionary employed in the Dictionary Learning-based scheme outperforms the SR-based baseline in two key aspects: (i) enhanced performance on small-scale datasets, and (ii) the ability to generate lower-dimensional yet effective representations. For brevity, we denote them as DL-based and SR-based.

\subsubsection*{Performance on Small-scale Dataset}

\begin{figure}[htbp]
\centering
\begin{minipage}[h]{0.5\textwidth}
\centering
\includegraphics[width=1\textwidth,height=0.35\textwidth]{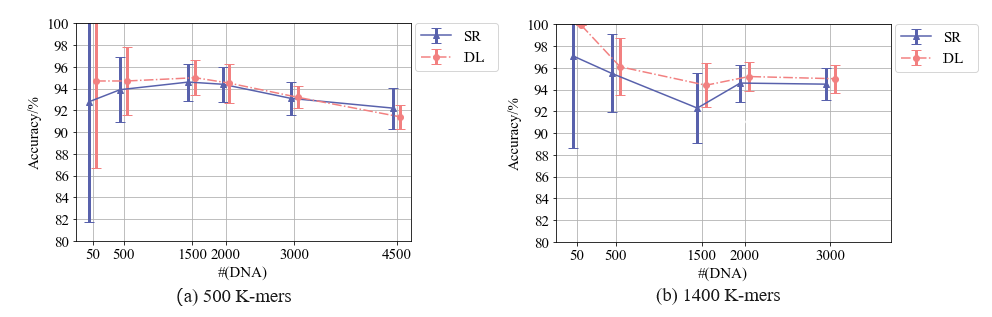}
\caption*{Subfigure (a) illustrates the classification accuracy on the first task based on the representation learned from different sizes of the dataset when the dictionary consists of 500 dymers, while subfigure (b) changes the basis/dictionary size to 1,400 dymers.}
\caption{An illustration of classification accuracy across different dataset sizes of two proposed schemes}
\label{6.4}
\end{minipage}
\end{figure}

We learned DNA representation under both schemes across a range of datasets, with only the dataset size varying. The obtained representations were then assessed the accuracy of the first classification task via 10-fold cross-validation. To evaluate generalizability, experiments were conducted for varying dictionary sizes. The results are shown in Figure  \ref{6.4}.

Key observations from Figure \ref{6.4} include:

\begin{enumerate}[i.]
    \item As the dataset size increases, the complexity of the classification tasks grows, leading to a slight decrease in accuracy for both methods.
    \item When the dictionary size is small (e.g., 500), the DL-based method yields higher accuracy than the SR-based method on smaller datasets, though the advantage diminishes with increasing dataset size.
\item When the dictionary size is large (e.g., 1,400), the DL-based method continues to surpass the SR-based method, particularly on smaller datasets.
\end{enumerate}

These findings suggest that the DL-based scheme is especially effective on small-scale datasets, mitigating data scarcity. The learnable dictionary better captures frequent, task-specific dymers for the given DNA sequences. In contrast, the SR-based approach tends toward a more generic or imbalanced distribution on small-scale datasets, making it difficult to identify representative dymers and thereby limiting expressiveness. 


\subsubsection*{Performance on Lower-Dimensional Representation}

\begin{figure}[htbp]
\centering
\begin{minipage}[h]{0.5\textwidth}
\centering
\includegraphics[width=0.75\textwidth,height=0.4\textwidth]{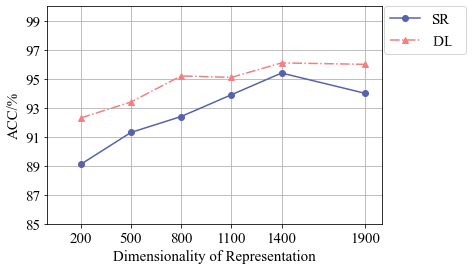}
\caption{An illustration of classification accuracy across different dimensionalities of representation from two proposed schemes}
\label{6.5}
\end{minipage}
\end{figure}

Figure \ref{6.5} presents the accuracy of the first classification task as a function of the dimensionality of the learned representations, comparing the DL-based and SR-based schemes trained on the same dataset. The results indicate that the DL-based scheme yields significantly lower-dimensional representations while achieving comparable or superior performance relative to the SR-based scheme. For example, an 800-dimensional representation learned by the DL-based approach attains performance competitive with, or better than, a 1,900-dimensional representation produced by the sparse representation learning (SR) baseline.


Except the promoter classification, our representation scheme could benefit plenty of DNA downstream applications.

\subsection{DNA Clustering and Phylogenetic Tree}

DNA, as the carrier of evolution-related biological information, elucidates relationships among species through similarities. Our representation scheme reconstructs the DNA sequence with dymers exhibiting distinct biochemical meanings. As a result, it facilitates the hierarchical clustering of DNA sequences and enables the generation of a phylogenetic tree, which illustrates the evolutionary relationships among different species.

\subsubsection{Dataset}

\textbf{The first exon of $\beta-$globin gene of 11 species}~\cite{JIN2017342}: includes Human, Goat, North American opossum, Gallus, Black lemur, House mouse, Rabbit, Norway rat, Gorilla, Bovine, and Chimpanzee. This dataset is commonly used in DNA similarity analysis.

\subsubsection{Metrics}

According to ~\cite{CK}, the evaluation of the accuracy of a phylogenetic tree can be compared with authoritative ones or results from other independent methods.

Qualitative standards derived from previous studies dictate that: (1) Primates ought to be grouped as closely related as possible; (2) Rodents should form a cohesive cluster; (3) All mammals should unite to form a single cluster; and (4) Non-mammalian species should be clearly distinguished from mammals.

\subsubsection{Experiments}

We compute the similarity between $m$ different DNA representations $\mathcal{A}^*_{:d_{i}:}$, and obtain the distance matrix $\textbf{Dis} \in \textbf{R}^{m*m}$. Here, we use the cosine similarity. Then, hierarchical clustering is conducted based on the distance matrix $\textbf{Dis}$, and the resulting clusters are visualized as a phylogenetic tree.

\begin{figure}[htbp]
\begin{minipage}[t]{\linewidth}
\centering
\includegraphics[width=1\textwidth,height=0.5\textwidth]{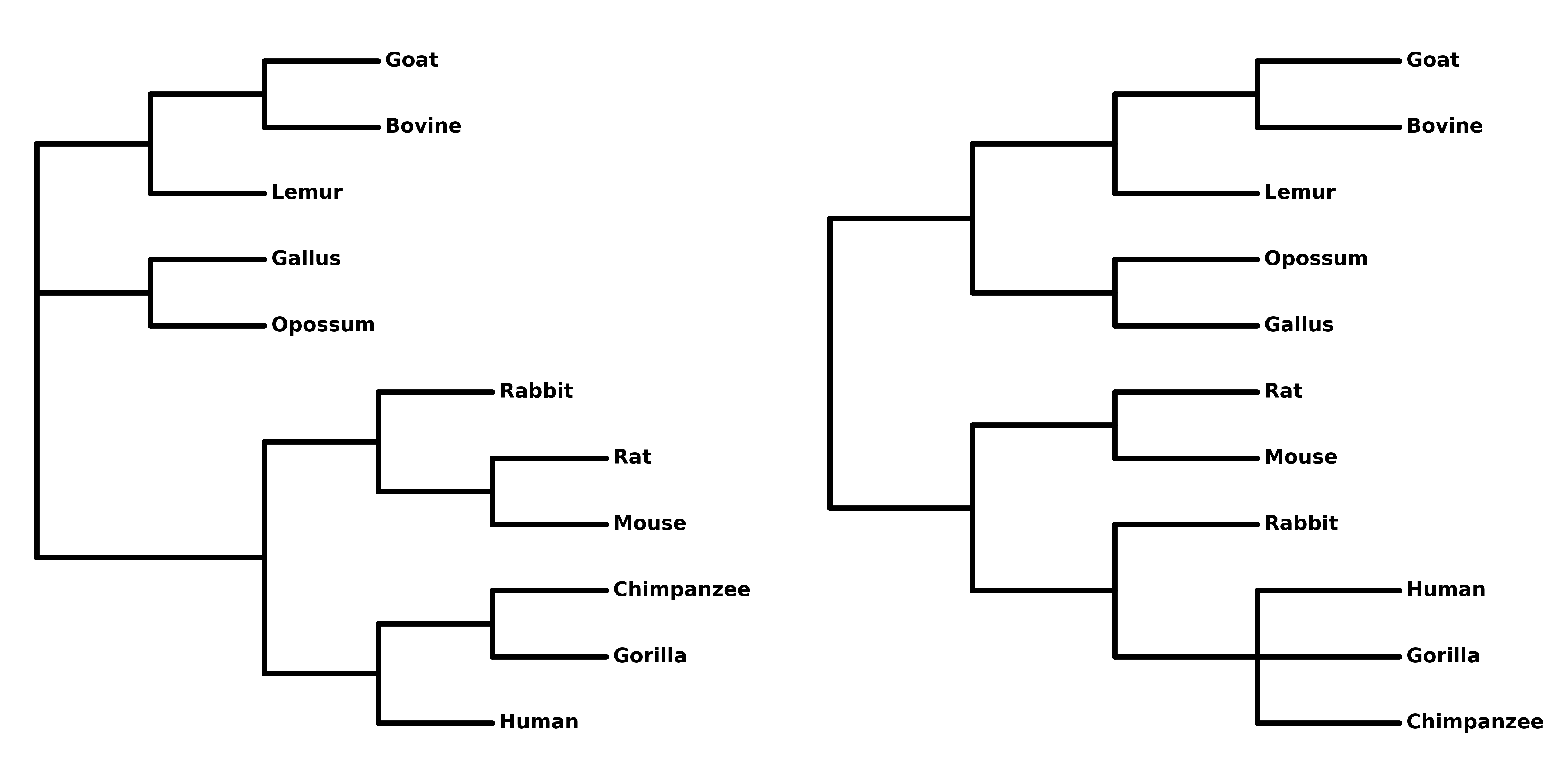}
\captionsetup{font={scriptsize}}
\caption{The left one is generated based on our DNA encoding scheme, while the right one is the results based on the multiple alignment CLUSTAL W}
\end{minipage}
\label{5.68}
\end{figure}

\subsubsection{Results}

Our phylogenetic tree is shown in the left image of Figure 9, while the right image depicts the results based on the multiple alignment method CLUSTAL W. It is obvious that our result clusters three primates, Human, Gorilla, and Chimpanzee together. Non-mammals, Opossum and Gallus, are distinct from mammals and form an independent cluster. Two rodents, Mouse and Rat, are categorized into one cluster. Goat and Bovine, both herbivores, are within the same cluster.

Analyzing the top 10 dymers of the highest likelihood for each species allows us to interpret the clustering results further. For primate, 6 out of 10 dymers are shared. Especially for human and gorilla, 8 out of 10 dymers are the same. Bovine and Goat have almost the same top10 list, similar to Rat and Mouse. Non-mammals, like Opossum and Gallus, are quite different from mammals. For example, \textit{ACGCCG} and \textit{CGCGCC} are especially frequent for Opossum and Gallus compared to mammals. \textit{CCAG}, \textit{TTAC} and \textit{TGAAA} are mostly appears in all species' DNA sequences.

\subsection{Motif Detection}

Motifs encapsulate the compositional patterns of dymers serving specific functions, such as binding sites. Typically, motifs are short patterns, around 9 base pairs in length, represented by position weight matrices.

In practical applications, researchers often conduct enrichment analyses on all DNA sequences to identify frequent dymers as instances of motifs. These identified instances are then summarized together to obtain a motif. Similarly, the dymer dictionary $\textbf{D}^*$ comprises those frequent dymers with high utilization. dymers in $\textbf{D}^*$ can be regarded as instances of motifs due to their significant enrichment compared to others in $\textbf{D}$.

\subsubsection{Dataset Description}

\textbf{Transcribed Pseudogenes in Homo Sapiens}: We collect 100 DNA sequences of homo sapiens (GRCh38.p14) from \textit{NCBI} datasets from National Library of Medicine. These 100 DNA sequences are all transcribed pseudogenes, with lengths ranging from 100 to 140 base pairs.

\subsubsection{Metrics}

To evaluate the frequent dymers captured by our method, we compare them with validated motifs. Here's the procedure:

1. Select the topk K-mers with the highest likelihood from the dymer dictionary $\textbf{D}^*$. Since motifs are typically DNA segments of about 9 nucleotides, select K-mers that are not too long or too short.

2. Find the most similar motif of \textit{Homo sapiens} for each K-mer based on the given score in \textit{JASPAR}, a prestigious open-access database of curated, non-redundant transcription factor binding profiles for multiple species.

3. Conduct alignment between the identified K-mers and the corresponding motifs. Fill in gaps within the sequence to maximize the number of matched nucleotides and record the number of gaps $g_{\phi_j}$ utilized for each K-mer $\phi_j$.

4. Compute the average number of filled gaps based on all top-q K-mers $\frac{\sum_{j} g_{\phi_j}}{q}$. This average number of filled gaps for the topk K-mers demonstrates the difference between motif instances found by our method and those validated motifs. 

\subsubsection{Results}

\begin{table}[!h]
\centering
\caption{The average number of filled gaps for topk K-mers}
\setlength{\tabcolsep}{10mm}
\begin{tabular}{lc}
  \toprule
   & $Avg.\ \#(filled\ gaps)$ \\
  \hline
  Top10 & 0.1  \\
  Top50 & 0.3 \\
  Top100 & 0.36 \\
  \bottomrule
\end{tabular}
\label{tab5.3}
\end{table}

\begin{figure}[htbp]
\begin{minipage}[t]{\linewidth}
\centering
\includegraphics[width=1\textwidth,height=0.55\textwidth]{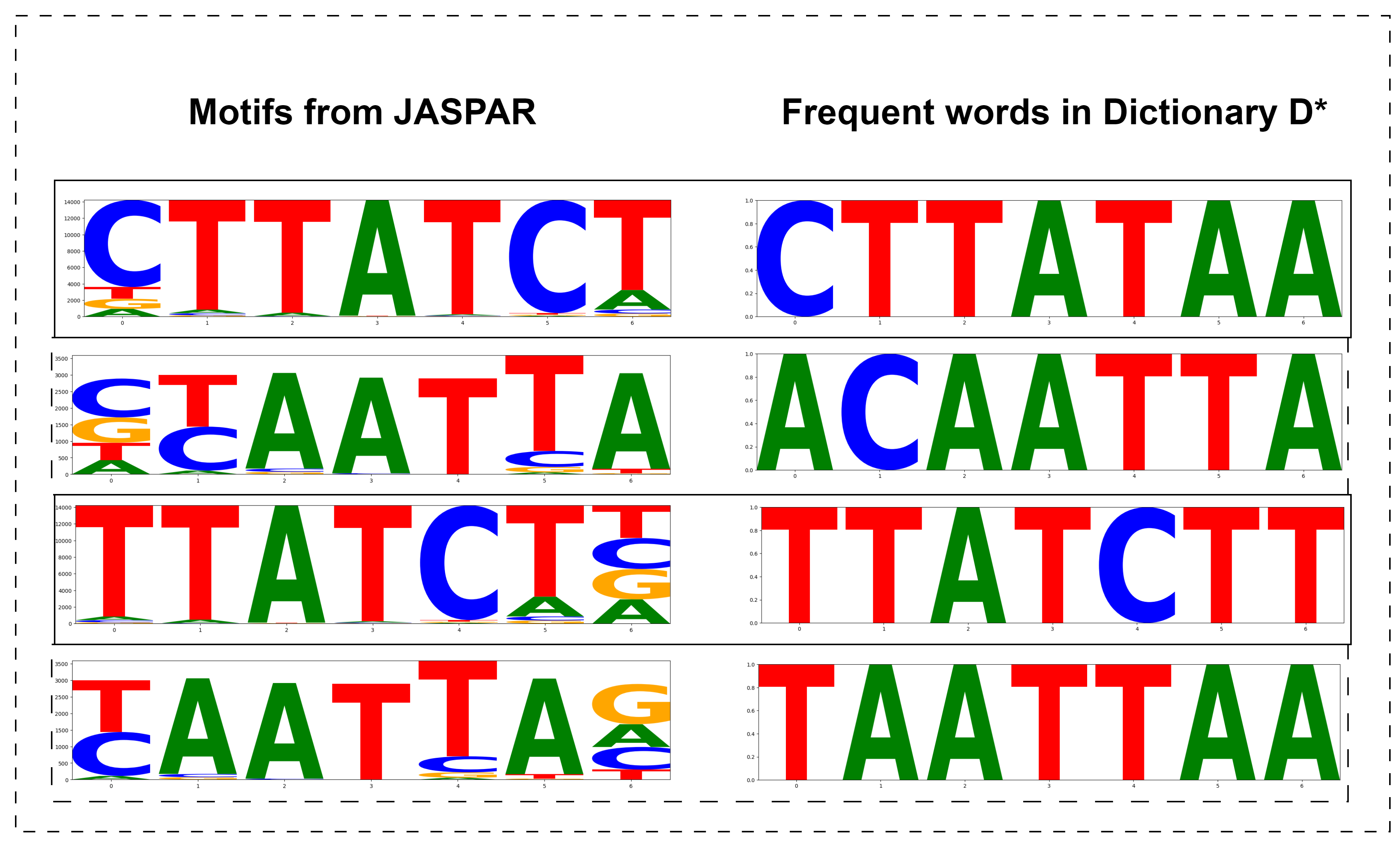}
\caption*{Many K-mers from dymer dictionary perfectly match existing and acknowledged motifs from homo species according to \textit{JASPAR}. The left column is from \textit{JASPAR} and the right column is from dymer dictionary.}
\caption{Results from motif detection}
\label{5.8}
\end{minipage}
\end{figure}

After analyzing the top10, top50, and top100 K-mers in Table \ref{tab5.3}, we find that the average number of mismatched nucleotides is less than 0.4. This indicates that, on average, less than one base per sequence is incorrect. In other words, the K-mers extracted by our method could potentially be valid instances of motifs and may help identify new, unseen motifs. Here are some examples illustrated in Figure \ref{5.8}.

\section{Conclusion}
In this study, we have proposed an innovative DNA sequence representation scheme, \textbf{Dy-mer}, inspired by dictionary learning, to address the limitations of existing methods in capturing the explainable features of DNA sequences. By formulating the problem as a tensor-based optimization, we successfully extract a frequent dymer dictionary comprising dynamic-length K-mers, enabling the reconstruction of DNA sequences through simple concatenation. Dy-mer offers a transparent framework for interpreting biological information encoded in DNA, ensuring both robustness and explainability in DNA representation.

Through extensive experiments, we have demonstrated the effectiveness of Dy-mer in various biological applications. Our approach outperforms existing state-of-the-art methods in DNA promoter classification. Furthermore, Dy-mer proves effective in DNA clustering and motif detection, providing rich interpretability essential for advancing biological research.

The success of Dy-mer underscores the importance of developing innovative and explainable DNA representation schemes to fully harness the potential of machine learning in deciphering biological data. By providing both robustness and interpretability, Dy-mer paves the way for advancements in understanding DNA sequences and their role in biological processes. In future endeavors, we aim to explore the application of our representation scheme on more biological tasks.

\section*{Acknowledgment}

This work is supported in part by the Natural Science Foundation of China (Grant 62371270)

\bibliographystyle{plainnat}
\bibliography{ref3} 

@article{xiao_ipsw2l-pseknc_2019,
  title={iPSW (2L)-PseKNC: A two-layer predictor for identifying promoters and their strength by hybrid features via pseudo K-tuple nucleotide composition},
  author={Xiao, Xuan and Xu, Zhao-Chun and Qiu, Wang-Ren and Wang, Peng and Ge, Hui-Ting and Chou, Kuo-Chen},
  journal={Genomics},
  volume={111},
  number={6},
  pages={1785--1793},
  year={2019},
  publisher={Elsevier}
}

@article{JIN2017342,
    title = {Similarity/dissimilarity calculation methods of DNA sequences: A survey},
    journal = {Journal of Molecular Graphics and Modelling},
    volume = {76},
    pages = {342-355},
    year = {2017},
    issn = {1093-3263},
    author = {Jin, Xin and Jiang, Qian and Chen, Yanyan and Lee, Shin-Jye and Nie, Rencan and Yao, Shaowen and Zhou, Dongming and He, Kangjian}
}

@article{noauthor_frontiers_nodate,
  title={Classifying promoters by interpreting the hidden information of DNA sequences via deep learning and combination of continuous fasttext N-grams},
  author={Le, Nguyen Quoc Khanh and Yapp, Edward Kien Yee and Nagasundaram, Nagarajan and Yeh, Hui-Yuan},
  journal={Frontiers in bioengineering and biotechnology},
  volume={7},
  pages={305},
  year={2019},
  publisher={Frontiers Media SA}
}

@inproceedings{zheng_deep_2021,
	address = {Nashville, TN, USA},
	title = {Deep {Convolutional} {Dictionary} {Learning} for {Image} {Denoising}},
	language = {en},
	urldate = {2023-12-24},
	booktitle = {2021 {IEEE}/{CVF} {Conference} on {Computer} {Vision} and {Pattern} {Recognition} ({CVPR})},
	publisher = {IEEE},
	author = {Zheng Hongyi and Yong Hongwei and Zhang Lei},
	month = jun,
	year = {2021},
	pages = {630--641},
}

@article{10.1093/nar/gkv1156,
    title={RegulonDB version 9.0: high-level integration of gene regulation, coexpression, motif clustering and beyond},
    author={Gama-Castro, Socorro and Salgado, Heladia and Santos-Zavaleta, Alberto and Ledezma-Tejeida, Daniela and Mu{\~n}iz-Rascado, Luis and Garc{\'\i}a-Sotelo, Jair Santiago and Alquicira-Hern{\'a}ndez, Kevin and Mart{\'\i}nez-Flores, Irma and Pannier, Lucia and Castro-Mondrag{\'o}n, Jaime Abraham and others},
    journal = {Nucleic Acids Research},
    volume = {44},
    number = {D1},
    pages = {D133-D143},
    year = {2015},
    month = {11},
    issn = {0305-1048}
}

@article{10.1093/nar/gku1019,
    author = {Lin, Hao and Deng, En-Ze and Ding, Hui and Chen, Wei and Chou, Kuo-Chen},
    title = "{iPro54-PseKNC: a sequence-based predictor for identifying sigma-54 promoters in prokaryote with pseudo k-tuple nucleotide composition}",
    journal = {Nucleic Acids Research},
    volume = {42},
    number = {21},
    pages = {12961-12972},
    year = {2014},
    month = {10},
}

@article{DEAVILAESILVA201422,
    title={DNA duplex stability as discriminative characteristic for Escherichia coli $\sigma$54-and $\sigma$28-dependent promoter sequences},
    author={e Silva, Scheila de Avila and Forte, Franciele and Sartor, Ivaine TS and Andrighetti, Tahila and Gerhardt, G{\"u}nther JL and Delamare, Ana Paula Longaray and Echeverrigaray, Sergio},
    journal = {Biologicals},
    volume = {42},
    number = {1},
    pages = {22-28},
    year = {2014}
}

@article{10.1093/bioinformatics/btx579,
    author = {Liu, Bin and Yang, Fan and Huang, De-Shuang and Chou, Kuo-Chen},
    title = "{iPromoter-2L: a two-layer predictor for identifying promoters and their types by multi-window-based PseKNC}",
    journal = {Bioinformatics},
    volume = {34},
    number = {1},
    pages = {33-40},
    year = {2017},
    month = {09},
}

@article{LI2006135,
    title={The recognition and prediction of $\sigma$70 promoters in Escherichia coli K-12},
  author={Li, Qian-Zhong and Lin, Hao},
    journal = {Journal of Theoretical Biology},
    volume = {242},
    number = {1},
    pages = {135-141},
    year = {2006}
}

@Article{biomimetics8020218,
AUTHOR = {Alakuş, Talha Burak},
TITLE = {A Novel Repetition Frequency-Based DNA Encoding Scheme to Predict Human and Mouse DNA Enhancers with Deep Learning},
JOURNAL = {Biomimetics},
VOLUME = {8},
YEAR = {2023},
NUMBER = {2},
ARTICLE-NUMBER = {218},
PubMedID = {37366813},
ISSN = {2313-7673},
}

@INPROCEEDINGS{5189632,
  author={Kwan, Hon Keung and Arniker, Swarna Bai},
  booktitle={2009 IEEE International Conference on Electro/Information Technology}, 
  title={Numerical representation of DNA sequences}, 
  year={2009},
  volume={},
  number={},
  pages={307-310},
}

@INPROCEEDINGS{950245,
  author={Cristea, P.},
  booktitle={Proceedings of the Sixth International Symposium on Signal Processing and its Applications (Cat.No.01EX467)}, 
  title={Genetic signal analysis}, 
  year={2001},
  volume={2},
  number={},
  pages={703-706 vol.2},
}

@article{AFREIXO201152,
title = {Genome analysis with distance to the nearest dissimilar nucleotide},
journal = {Journal of Theoretical Biology},
volume = {275},
number = {1},
pages = {52-58},
year = {2011},
author = {Afreixo, Vera and Bastos, Carlos AC and Pinho, Armando J and Garcia, Sara P and Ferreira, Paulo JSG}
}

@inproceedings{tang2023explainable,
  title={Explainable Trajectory Representation through Dictionary Learning},
  author={Tang, Yuanbo and Peng, Zhiyuan and Li, Yang},
  booktitle={Proceedings of the 31st ACM International Conference on Advances in Geographic Information Systems},
  pages={1--4},
  year={2023}
}

@article{BAI2011232,
title = {Similarity analysis of DNA sequences based on the EMD method},
journal = {Applied Mathematics Letters},
volume = {24},
number = {2},
pages = {232-237},
year = {2011},
author = {Fenglan Bai and Jihong Zhang and Junsheng Zheng}
}

@article{JIN2016325,
title = {A novel DNA sequence similarity calculation based on simplified pulse-coupled neural network and Huffman coding},
journal = {Physica A: Statistical Mechanics and its Applications},
volume = {461},
pages = {325-338},
year = {2016},
author = {Xin Jin and Rencan Nie and Dongming Zhou and Shaowen Yao and Yanyan Chen and Jiefu Yu and Quan Wang}
}

@article{CK,
title = {Position–Specific Statistical Model of DNA Sequences and Its Application for Similarity Analysis},
journal = {MATCH Communications in Mathematical and in Computer Chemistry},
volume = {73},
pages = {545-558},
year = {2015},
issn = {0340-6253},
author={Kuang, Chenkui and Liu, Xiaoqing and Wang, Junru and Yao, Yuhua and Dai, Qi},
}

@article{article,
author = {Song, Kai},
year = {2012},
month = {11},
pages = {963-971},
title = {Recognition of prokaryotic promoters based on a novel variable-window Z-curve method},
volume = {40},
journal = {Nucleic Acids Research}
}

@article{Sun2021iPTT2LA,
  title={iPTT (2 L)-CNN: A Two-Layer Predictor for Identifying Promoters and Their Types in Plant Genomes by Convolutional Neural Network},
  author={Sun, Ang and Xiao, Xuan and Xu, Zhaochun},
  journal={Computational and Mathematical Methods in Medicine},
  volume={2021},
  pages={1--9},
  year={2021},
  publisher={Hindawi Limited}
}

@article{doi:10.1021/acs.jcim.8b00368,
author = {Xue, Li and Tang, Bin and Chen, Wei and Luo, Jiesi},
title = {Prediction of CRISPR sgRNA Activity Using a Deep Convolutional Neural Network},
journal = {Journal of Chemical Information and Modeling},
volume = {59},
number = {1},
pages = {615-624},
year = {2019},
}

@article {PMID:34903170,
title={AttCRISPR: a spacetime interpretable model for prediction of sgRNA on-target activity},
author={Xiao, Li-Ming and Wan, Yun-Qi and Jiang, Zhen-Ran},
Number = {1},
Volume = {22},
pages={1--17},
Month = {December},
Year = {2021},
Journal = {BMC bioinformatics},
}

@article{SANTORSOLA20231,
title = {The promise of explainable deep learning for omics data analysis: Adding new discovery tools to AI},
journal = {New Biotechnology},
volume = {77},
pages = {1-11},
year = {2023},
author = {Mariangela Santorsola and Francesco Lescai}
}

@article{candes2006robust,
title={Robust uncertainty principles: Exact signal reconstruction from highly incomplete frequency information},
author={Cand{\`e}s, Emmanuel J and Romberg, Justin and Tao, Terence},
journal={IEEE Transactions on information theory},
volume={52},
number={2},
pages={489--509},
year={2006},
publisher={IEEE}
}

@article{donoho2006compressed,
title={Compressed sensing},
author={Donoho, David L},
journal={IEEE Transactions on information theory},
volume={52},
number={4},
pages={1289--1306},
year={2006},
publisher={IEEE}
}

@article{kmer1,
title={Genomic DNA k-mer spectra: models and modalities},
author={Benny Chor and David Horn and Nick Goldman and Yaron Levy and Tim Massingham},
journal={Genome biology},
volume={10},
number={10},
pages={R108},
year={2009},
}

@article{kmer2,
title={KAT: a K-mer analysis toolkit to quality control NGS datasets and genome assemblies},
author={Mapleson, Daniel and Garcia Accinelli, Gonzalo and Kettleborough, George and Wright, Jonathan and Clavijo, Bernardo J},
journal={Bioinformatics},
volume={33},
number={4},
pages={574-576},
year={2017},
}

@article{dna,
  title={Merriam-webster},
  author="{Dictionary, Merriam-Webster}",
  journal={On-line at http://www.mw.com/home.htm},
  volume={8},
  number={2},
  year={2002}
}

@article{ini,
title={DNA sequence requirements for transcriptional initiator activity in mammalian cells},
author={Javahery, Ramin and Khachi, Anita and Lo, Kiersten and Zenzie-Gregory, Beatrice and Smale, Stephen T},
journal={Mol Cell Biol},
volume={14},
number={1},
pages={116-127},
year={1994},
PMID={8264580},
PMCID={PMC358362}
}

@article{gcbox,
title={Importance of a flanking AT-rich region in target site recognition by the GC box-binding zinc finger protein MIG1},
author={Lundin, MARIA and Nehlin, Jan Olof and Ronne, Hans},
journal={Mol Cell Biol},
volume={14},
number={3},
pages={1979-1985},
year={1994},
PMID={8114729},
PMCID={PMC358557}
}

@article{tata,
title={The organization of the histone genes in Drosophila melanogaster: functional and evolutionary implications},
author={Lifton, RP and Goldberg, ML and Karp, RW and Hogness, DS},
journal={Cold Spring Harbor Symposia on Quantitative Biology},
volume={42},
number={2},
pages={1047-1051},
year={1978},
PMID={98262}
}

@article{cpg,
title={Cytosine methylation and CpG, TpG (CpA) and TpA frequencies},
author={Jabbari, Kamel and Bernardi, Giorgio},
journal={Gene},
volume={333},
pages={143-149},
year={2004},
PMID={15177689}
}

@article{caat,
title={The NF-YB/NF-YC Structure Gives Insight into DNA Binding and Transcription Regulation by CCAAT Factor NF-Y},
author={Romier, Christophe and Cocchiarella, Fabienne and Mantovani, Roberto and Moras, Dino},
journal={The Journal of Biological Chemistry},
volume={278},
number={2},
pages={1336-1345},
year={2002},
PMID={12401778}
}

@article{conseq,
title={Consensus sequence Zen},
author={Schneider, Thomas D},
journal={Applied Bioinformatics},
volume={1},
number={3},
pages={111-119},
year={2002},
PMID={15130839}
}

@article{motif,
title={What are DNA sequence motifs?},
author={Patrik D'haeseleer},
journal={Nat Biotechnol},
volume={24},
pages={423-425},
year={2006},
}

@article {Chu2022.11.06.515322,
author = {Shane Chu and Gary Stormo},
title = {Deep unfolded convolutional dictionary learning for motif discovery},
elocation-id = {2022.11.06.515322},
year = {2022},
publisher = {Cold Spring Harbor Laboratory},
eprint = {https://www.biorxiv.org/content/early/2022/11/10/2022.11.06.515322.full.pdf},
journal = {bioRxiv}
}

@article{ng2017dna2vec,
  title={dna2vec: Consistent vector representations of variable-length k-mers},
  author={Ng, Patrick},
  journal={arXiv preprint arXiv:1701.06279},
  year={2017}
}

@misc{NCBI,
    title = {Bethesda (MD): National Library of Medicine (US), National Center for Biotechnology Information},
    urldate = {April 3, 2024},
    year = {1988},
}

@article{ATT1,
author = {Sun, Ang and Xuan, Xiao and Xu, Zhaochun},
year = {2021},
month = {01},
pages = {1-9},
title = "{iPTT(2L)-CNN: A Two-Layer Predictor for Identifying Promoters and Their Types in Plant Genomes by Convolutional Neural Network}",
volume = {2021},
journal = {Computational and Mathematical Methods in Medicine},
}

@article{RANDIC2003202,
title = {Analysis of similarity/dissimilarity of DNA sequences based on novel 2-D graphical representation},
journal = {Chemical Physics Letters},
volume = {371},
number = {1},
pages = {202-207},
year = {2003},
author = {Milan Randić and Marjan Vračko and Nella Lerš and Dejan Plavšić}
}

@article{RANDIC20031,
title = {Novel 2-D graphical representation of DNA sequences and their numerical characterization},
journal = {Chemical Physics Letters},
volume = {368},
number = {1},
pages = {1-6},
year = {2003},
author = {Randi{\'c}, Milan and Vra{\v{c}}ko, Marjan and Ler{\v{s}}, Nella and Plav{\v{s}}i{\'c}, Dejan}
}

@article{BAI2007282,
title = {A representation of DNA primary sequences by random walk},
journal = {Mathematical Biosciences},
volume = {209},
number = {1},
pages = {282-291},
year = {2007},
author = {Fenglan Bai and Yingzhao Liu and Tianming Wang}
}

@misc{jaspar,
title = {JASPAR: an open-access database for eukaryotic transcription factor binding profiles Nucleic Acids Res}, 
year = {2004},
author={Sandelin, Albin and Alkema, Wynand and Engstr{\"o}m, P{\"a}r and Wasserman, Wyeth W and Lenhard, Boris}
}

@INPROCEEDINGS{6909690,
  author={Badri, Hicham and Yahia, Hussein and Aboutajdine, Driss},
  booktitle={2014 IEEE Conference on Computer Vision and Pattern Recognition}, 
  title={Robust Surface Reconstruction via Triple Sparsity}, 
  year={2014},
  volume={},
  number={},
  pages={2291-2298},
}

@article{castro2023exploring,
  title={Exploring the Effect of Sparse Recovery on the Quality of Image Superresolution},
  author={Castro, Antonio},
  journal={arXiv preprint arXiv:2308.02714},
  year={2023}
}

@ARTICLE{7102696,
  author={Zhang Zheng and Xu Yong and Yang Jian et al.},
  journal={IEEE Access}, 
  title={A Survey of Sparse Representation: Algorithms and Applications}, 
  year={2015},
  volume={3},
  number={},
  pages={490-530},
}

@ARTICLE{7903603,
  author={Xu, Yong and Li, Zhengming and Yang, Jian and Zhang, David},
  journal={IEEE Access}, 
  title={A Survey of Dictionary Learning Algorithms for Face Recognition}, 
  year={2017},
  volume={5},
  number={},
  pages={8502-8514},
}

@ARTICLE{coor,
  author={Wright, Stephen J},
  journal={Mathematical Programming}, 
  title={Coordinate descent algorithms}, 
  year={2015},
  volume={151},
  number={1},
  pages={3-34},
}

@book{coor2,
  title={Convex optimization},
  author={Boyd, Stephen P and Vandenberghe, Lieven},
  year={2004},
  publisher={Cambridge university press}
}

@article{cgd,
  title={Efficiency of coordinate descent methods on huge-scale optimization problems},
  author={Nesterov, Yu},
  journal={SIAM Journal on Optimization},
  volume={22},
  number={2},
  pages={341--362},
  year={2012},
  publisher={SIAM}
}

@article{cgd2,
  title={A coordinate gradient descent method for nonsmooth separable minimization},
  author={Tseng, Paul and Yun, Sangwoon},
  journal={Mathematical Programming},
  volume={117},
  pages={387--423},
  year={2009},
  publisher={Springer}
}

@article{llm,
author = {Wang Tao, Luo Zeyu.},
title = {Large language models transform biological research: from architecture to utilization},
journal = {Science China Information Sciences},
volume = {68},
issue = {7},
pages = {170101},
year = {2025}
}

@article{nature,
author = {Sanabria Melissa, Hirsch Jonas, Joubert Pierre M.},
title = {DNA language model GROVER learns sequence context in the human genome},
journal = {Nature Machine Intelligence},
volume = {6},
issue = {8},
pages = {911-923},
year = {2024}
}

@inproceedings{
zhou2024dnabert,
title={{DNABERT}-2: Efficient Foundation Model and Benchmark For Multi-Species Genomes},
author={Zhihan Zhou and Yanrong Ji and Weijian Li and Pratik Dutta and Ramana V Davuluri and Han Liu},
booktitle={The Twelfth International Conference on Learning Representations},
year={2024},
}

@article{DBLP:journals/corr/abs-2307-05628,
  publtype={informal},
  author={Daoan Zhang and Weitong Zhang and Bing He and Jianguo Zhang and Chenchen Qin and Jianhua Yao},
  title={DNAGPT: A Generalized Pretrained Tool for Multiple DNA Sequence Analysis Tasks},
  year={2023},
  cdate={1672531200000},
  journal={CoRR},
  volume={abs/2307.05628},
}

@article{10.1093/bioinformatics/btab083,
    author = {Ji, Yanrong and Zhou, Zhihan and Liu, Han and Davuluri, Ramana V},
    title = {DNABERT: pre-trained Bidirectional Encoder Representations from Transformers model for DNA-language in genome},
    journal = {Bioinformatics},
    volume = {37},
    number = {15},
    pages = {2112-2120},
    year = {2021},
    month = {02},
}

@inproceedings{10.5555/3666122.3667994,
author = {Nguyen, Eric and Poli, Michael and Faizi, Marjan and Thomas, Armin W. and Sykes, Callum Birch and Wornow, Michael and Patel, Aman and Rabideau, Clayton and Massaroli, Stefano and Bengio, Yoshua and Ermon, Stefano and Baccus, Stephen A. and R\'{e}, Christopher},
title = {HyenaDNA: long-range genomic sequence modeling at single nucleotide resolution},
year = {2023},
publisher = {Curran Associates Inc.},
address = {Red Hook, NY, USA},
booktitle = {Proceedings of the 37th International Conference on Neural Information Processing Systems},
articleno = {1872},
numpages = {25},
location = {New Orleans, LA, USA},
series = {NIPS '23}
}

@article{sanabria2024dna,
title={DNA language model GROVER learns sequence context in the human genome},
author={Sanabria, Melissa and Hirsch, Jonas and Joubert, Pierre M and Poetsch, Anna R},
journal={Nature Machine Intelligence},
volume={6},
number={8},
pages={911--923},
year={2024},
publisher={Nature Publishing Group UK London}
}

@inproceedings{arora_simple_2015,
	title = {Simple, {Efficient}, and {Neural} {Algorithms} for {Sparse} {Coding}},
	booktitle = {Proceedings of {The} 28th {Conference} on {Learning} {Theory}},
	publisher = {PMLR},
	author = {Arora, Sanjeev and Ge, Rong and Ma, Tengyu and Moitra, Ankur},
	month = jun,
	year = {2015},
	note = {ISSN: 1938-7228},
	pages = {113--149},
}

@article{dicmotif,
    author = {Chu, Shane K and Stormo, Gary D},
    title = {Finding motifs using DNA images derived from sparse representations},
    journal = {Bioinformatics},
    volume = {39},
    number = {6},
    pages = {btad378},
    year = {2023},
    month = {06}
}

@article{common,
  title={Independent component analysis, A new concept?},
  author={Comon, Pierre },
  journal={Signal Processing},
  volume={36},
  number={3},
  pages={287-314},
  year={1994}
}

@ARTICLE{Wang,
  author={Kuansan Wang and Chin-Hui Lee and Biing-Hwang Juang},
  journal={IEEE Signal Processing Letters}, 
  title={Selective feature extraction via signal decomposition}, 
  year={1997},
  volume={4},
  number={1},
  pages={8-11}
}

@ARTICLE{789,
  author={M S Lewicki, T J Sejnowski},
  journal={Neural Comput}, 
  title={Learning overcomplete representations}, 
  year={2000},
  volume={12},
  issue={2},
  pages={337-65}
}

@ARTICLE{field,
  author={Field, David J.},
  journal={Neural Computation}, 
  title={What Is the Goal of Sensory Coding?}, 
  year={1994},
  volume={6},
  number={4},
  pages={559-601}
}

@article{WATANABE1981381,
author = {Satosi Watanabe},
title = {Pattern recognition as a quest for minimum entropy},
journal = {Pattern Recognition},
volume = {13},
number = {5},
pages = {381-387},
year = {1981}
}

@article{survey1,
author = {Kenneth Kreutz-Delgado, Joseph F Murray, Bhaskar D Rao},
title = {Dictionary learning algorithms for sparse representation},
journal = {Neural Comput},
volume = {15},
issue = {2},
pages = {349-96},
year = {2003}
}

@article{survey2,
author = {Zhao Rongchang,Li Hong,Liu Xiyao},
title = {A Survey of Dictionary Learning in Medical Image Analysis and Its Application for Glaucoma Diagnosis},
journal = {Archives of Computational Methods in Engineering},
volume = {28},
issue = {2},
pages = {463-471},
year = {2021}
}

@ARTICLE{258082,
  author={Mallat, S.G. and Zhifeng Zhang},
  journal={IEEE Transactions on Signal Processing}, 
  title={Matching pursuits with time-frequency dictionaries}, 
  year={1993},
  volume={41},
  number={12},
  pages={3397-3415},
  doi={10.1109/78.258082}}

\appendix

\section{Basic Operations in Tensor Notation}

Here are definitions for several basic operations in Tensor notion used in our formulation.

\begin{myDef}
    Kronecker Product. 
    
    Given 2 tensors $A\in \mathbf{R}^{I\times J}$ and $B\in \mathbf{R}^{K\times L}$, the Kronecker product of these two tensors is a tensor $K \in \mathbf{(IK)\times (JL)}$. Use $\otimes$ to represent Kronecker product.

    $$
    K = A \otimes B = 
    \begin{bmatrix} 
            a_{00}B & a_{01}B & \dots & a_{0J}B \\ 
            a_{10}B & a_{11}B & \dots & a_{1J}B \\ 
            \dots & \dots & \dots & \dots \\ 
            a_{I0}B & a_{01}B & \dots & a_{IJ}B 
        \end{bmatrix} 
    $$

    Equivalent to saying, let every element in $A$ times $B$.
    
\end{myDef}

\begin{myDef}
    Given a tensor $A\in \mathbf{R}^{I\times J\times L}$ and a matrix $B\in \mathbf{R}^{(JL)\times K}$. The product between tensor $A$ and matrix $B$ is defined as $A \times_{i} B$, which means unfold the tensor into a matrix through $i_{th}$ dimension, namely mode-$i$ Matricization $A_{(1)}$, then multiply them to get the result.

    For instance,

    $$
    \begin{aligned}
        P & = A \times_{(1)} B \\
        & = 
            \begin{bmatrix}
                \\
                \begin{bmatrix}
                1 & 2 \\ 
                3 & 4 \\ 
                \end{bmatrix} \\
                \\
                \begin{bmatrix} 
                5 & 6 \\ 
                7 & 8 \\ 
                \end{bmatrix}
                \\
                \\
            \end{bmatrix} \times_{1} 
            \begin{bmatrix} 
                a & b \\ 
                c & d \\ 
            \end{bmatrix} \\
        \Leftrightarrow \ P_{(1)} & = BA_{(1)} \\
            & = \begin{bmatrix} 
                a & b \\ 
                c & d \\ 
                \end{bmatrix}
                \begin{bmatrix} 
                1 & 2 & 5 & 6 \\ 
                3 & 4 & 7 & 8 \\ 
                \end{bmatrix}  \\
            & = \begin{bmatrix} 
                a+3b & 2a+4b & 5a+7b & 6a+8b \\ 
                c+3d & 2c+4d & 5c+7d & 6c+8d \\ 
                \end{bmatrix} \\
        P & = \begin{bmatrix} 
                \\
                \begin{bmatrix} 
                a+3b & 2a+4b \\ 
                5a+7b & 6a+8b \\ 
                \end{bmatrix} \\
                \\
                \begin{bmatrix} 
                c+3d & 2c+4d \\ 
                5c+7d & 6c+8d \\
                \end{bmatrix}
                \\
                \\
            \end{bmatrix}        
    \end{aligned}
    $$
    
    Beware that the mode-$i$ Matricization $A_{(1)}$ should be aligned with the given matrix $B$ on the dimension.
    
\end{myDef}

\section{Algorithm Pseudocode}
Algorithm~\ref{alg:dymer_dl} details the optimization process for learning sparse representations of DNA sequences within the Dy-mer framework. The algorithm employs an alternating optimization strategy, iteratively updating the assignment tensor $\mathcal{A}$ and the dictionary tensor $\mathcal{D}$ to minimize a composite objective function.

In each iteration, the process consists of two main steps:
\begin{enumerate}
    \item \textbf{Sparse Representation Update:} With the dictionary $\mathcal{D}$ held fixed, the assignment tensor $\mathcal{A}$ is updated by minimizing a loss function comprising a reconstruction loss, a dictionary complexity term ($\mathcal{L}_{DC}$), and an $L_1$ regularization term.
    
    \item \textbf{Dictionary Update:} Subsequently, with the newly updated assignment tensor $\mathcal{A}$ held fixed, the dictionary $\mathcal{D}$ is updated to minimize the reconstruction error between the original DNA sequences and the sequences reconstructed from $\mathcal{A}$ and $\mathcal{D}$.
\end{enumerate}
These two steps are repeated for a predefined number of epochs ($T_{max}$), yielding the final optimized dictionary $\mathcal{D}^*$ and the corresponding sparse DNA representations $\mathcal{A}^*$.

The time complexity is $O(T_{max} \cdot m \cdot n \cdot L_2 \cdot L_3)$, and the space complexity is $O(m \cdot L_1 + n \cdot L_2 + n \cdot m \cdot L_3)$. The time complexity is determined by the main loop that runs for $T_{max}$ epochs. Within each epoch, the computational bottleneck is the two update steps (for the assignment tensor A and the dictionary D). Both steps require calculating gradients based on the reconstruction loss, which is dominated by a convolutional operation ($A \odot D$) across all data. Specifically, m is the number of input DNA sequences, n is the number of dymers in the dictionary, $L_2$ is the length of a dymer, and $L_3$ is the length of the activation map.

The space complexity is dictated by the memory required to store the primary tensors: the input DNA sequence tensor X of size $m \times 4 \times L_{1}$, the dymer dictionary D of size $n \times 4 \times L_{2}$, and the assignment tensor A of size $n \times m \times L_{3}$.

\section{Dymer Distribution over dymer length}

In this section, the distribution of different dymer lengths is plotted as a histogram as follows. It is illustrated that the learned dymers vary from 3 to 7.

\begin{figure}[htbp]
\begin{minipage}[t]{\linewidth}
\centering
\includegraphics[width=0.8\textwidth,height=0.4\textwidth]{figures/kmer_distribution_discrete_bar_fixed.png}
\captionsetup{font={scriptsize}}
\caption{The distribution of dymer lengths}
\end{minipage}
\label{5.68}
\end{figure}

\end{document}